\documentclass{aa}  

\pdfoutput=1

\usepackage[hidelinks]{hyperref}

\renewcommand{\rm}{\mathrm}
\def\inv{^{-1}}
\def\({\left(}
\def\r){\right)}
\def\b#1{\bm{#1}}

\def\Dds{D_{\rm{ds}}}
\def\Ds{D_{\rm{s}}}
\def\la{\left<}
\def\ra{\right>}
\def\gammao{\gamma^{\text{obs}}}
\def\gammaoh{\hat{\gamma}^{\text{obs}}}

\def\d{\rm{d}}
\def\bthp{\b \theta' + \b \theta}
\def\bth{\b \theta'}
\def\i{\rm{i}}

\usepackage{bm}
\usepackage{verbatim}
\usepackage{xcolor}
\usepackage{graphicx}
\usepackage{txfonts}
\usepackage[section]{placeins}

\usepackage{todonotes}

\begin{document}

   \title{The effects of varying depth in cosmic shear surveys}


   \author{Sven Heydenreich\inst{1}
          \and Peter Schneider\inst{1} 
          \and Hendrik Hildebrandt\inst{2,1}
          \and Marika Asgari\inst{3}
          \and Catherine Heymans\inst{3,2} 
          \and Benjamin Joachimi\inst{4}
          \and Konrad Kuijken\inst{5}
          \and Chieh-An Lin\inst{3}
          \and Tilman Tr\"{o}ster\inst{3}
          \and Jan Luca van den Busch\inst{2,1}
          }
   \institute{Argelander-Institut f\"ur Astronomie, Auf dem H\"ugel 71, 53121 Bonn, Germany 
\and 
The German Centre for Cosmological Lensing, Astronomisches Institut, Ruhr-Universit\"at Bochum, Universit\"atsstr. 150, 44801 Bochum, Germany 
\and
Institute for Astronomy, University of Edinburgh, Royal Observatory, Blackford Hill, Edinburgh EH9 3HJ, UK
\and
Department of Physics and Astronomy, University College London, Gower Street, London WC1E 6BT, UK
\and
Leiden Observatory, Leiden University, Niels Bohrweg 2, 2333 CA Leiden, The Netherlands
\\ \email{sven@astro.uni-bonn.de, peter@astro.uni-bonn.de}
             }

   \date{Received 21 October 2019; accepted 02 December 2019}

  \abstract
   {We present a semi-analytic model for the shear two-point correlation function of a cosmic shear survey with non-uniform depth. Ground-based surveys are subject to depth variations that primarily arise through varying atmospheric conditions. For a survey like the Kilo-Degree Survey (KiDS), we find that the measured depth variation increases the amplitude of the observed shear correlation function at the level of a few percent out to degree-scales, relative to the assumed uniform-depth case. The impact on the inferred cosmological parameters is shown to be insignificant for a KiDS-like survey. For next-generation cosmic shear experiments, however, we conclude that variable depth should be accounted for.}

   \maketitle

%
\section{Introduction}
The discovery of cosmic shear has provided us with a new and powerful cosmological tool to empirically test the standard model of cosmology and to determine its parameters. Contrary to the analysis of the cosmic microwave background \citep[CMB, e.g.,~by][]{2018arXiv180706209P}, cosmic shear is more sensitive to the properties of the low-redshift large-scale structure and, thus, provides an excellent consistency check for the standard model. Current cosmic shear surveys are particularly sensitive to the parameter $S_8=\sigma_8 \sqrt{\Omega_{\rm m}/0.3}$, where $\sigma_8$ characterizes the normalization of the matter power spectrum and $\Omega_{\rm m}$ is the matter density parameter. Constraints on $S_8$ from the three current major cosmic shear results are all consistent with the CMB analysis by \citet{2018arXiv180706209P}. It is interesting to note, however, that they all favor values that are slightly lower than the Planck constraints of $S_8 = 0.830 \pm 0.013$. \citet{2018arXiv180909148H} report $S_8 = 0.800^{+0.029}_{-0.028}$ from an analysis of the Subaru Hyper Suprime-Cam survey, \citet[][hereafter H18]{2018arXiv181206076H} obtained $0.737_{-0.036}^{+0.040}$ from KiDS+VIKING data, and \citet{2018PhRvD..98d3528T} constrain $S_8=0.782\pm 0.027$ using the Dark Energy Survey (DES).  Combined analyses of DES and KiDS data \citep{Joudaki:2019, asgari/etal:2019} result in a $\sim 3 \sigma$ tension with the CMB value for $S_8$. If this tension is not the manifestation of an unaccounted systematic effect, in either the cosmic shear surveys \citep{Mandelbaum:2019} or the Planck mission \citep{2016ApJ...818..132A}, it certainly merits attention. It could be interpreted as a sign of new physics exemplified by massive neutrinos \citep{2014PhRvL.112e1303B}, time-varying dark energy, or modified gravity \citep{2016A&A...594A..14P} and coupling within the dark sector \citep{Kumar:2019}.  It could also, however, prove to be a simple statistical coincidence.


For current cosmic shear surveys, the estimated systematic error is becoming comparable in magnitude to the statistical error, implying that for next-generation surveys, a significant reduction of systematic errors is necessary. With surveys like the Large Synoptic Survey Telescope \citep[LSST,][]{Ivezic:2008} and Euclid \citep{Laureijs:2011} soon to begin, systematic effects in gravitational lensing have received a large amount of attention \citep[see][and references therein]{Mandelbaum:2019}.

In this paper, we focus on systematic effects induced by variation in survey depth that is so far unaccounted for in cosmic shear analyses \citep{Vale:2004}.   For a survey with a fixed exposure time, varying atmospheric conditions, dithering strategies and galactic extinction all contribute to an inhomogeneous limiting magnitude as a function of sky position.   In order to assess the impact of variable depth for current and future surveys, we build an analytical model for the effect based on the survey specifications of the Kilo-Degree Survey \citep[KiDS,][]{Kuijken:2015}.  To the first order, the depth variation in KiDS can be modeled by a piece-wise constant depth function, which varies between each $1\,\text{deg}^2$ square pointing.   KiDS object detection is defined in the $r$-band as these images were chosen to be significantly deeper in comparison to the other optical and near-infra red filters.   We, therefore, quantify survey depth with the limiting $r$-band magnitude, as defined in \citet{2017A&A...604A.134D}.  We defer the study of multi-band variable depth and its impact on photometric redshift accuracy for a future work.

This work is complementary to the analysis of \citet{Guzik:2005}, who investigate the effect of a general, position-dependent multiplicative shear bias on the shear power spectrum.  In principle, the varying depth of the source galaxy sample that we study here could be recast as a varying effective shear bias.  The inclusion of an inhomogeneous distribution of source galaxies has also been explored using mock catalogs of the Subaru Hyper Suprime-Cam Survey \citep{Shirasaki:2019}, with a focus on resulting estimation of the cosmic shear covariance matrix.

%

In Sect. \ref{sec:analytic_models}, we will introduce two simple toy models to understand this effect and analyze the impact on the cosmic shear power spectrum. In Sect. \ref{sec:xipm_semianalytic}, we will estimate the effect on the shear correlation functions $\xi_\pm$ using a semi-analytic model. We will present our results in Sect. \ref{sec:results}. In Sect. \ref{sec:discussion}, we will discuss our results and comment on the impact of our used simplifications. In the appendices, we present the full derivation of our model for finite field surveys.
We assume the standard weak gravitational lensing formalism, a summary of which can be found in \citet{2001PhR...340..291B}.

\section{Simple, analytic toy models}
\label{sec:analytic_models}
For our first analysis, we assume that all the matter between the sources and observer is concentrated in a single lens plane of distance $D_{\rm d}$ from the observer. If we then distribute sources at varying distances $D_{\rm s}$, two effects become apparent: firstly, the lensing efficiency $D_{\rm{ds}}/D_{\rm s}$ varies, where $D_{\rm{ds}}$ is the distance between the lens plane and the respective source. Secondly, and more importantly, for a more distant source, more matter is concentrated between the source and the observer, leading to a stronger shear signal.

Assuming that the depth and, thus, the source redshift population, only varies between pointings of the camera, an observer will measure a shear signal that is modified by a step-like depth-function, $\gammao(\b\theta)=W(\b \theta)\gamma(\b\theta)$, where $W$ is proportional to the mean of the lensing efficiency $\Dds/\Ds$ of one pointing and $\gamma$ denotes the shear that this pointing would experience if it were of the average depth.
We can parametrize $W$ as $W(\bm{\theta}) = 1+w(\bm{\theta})$. This implies that $\langle w(\bm{\theta})\rangle=0$ holds, where $\langle \cdot\rangle$ denotes the average over all pointings.

\subsection{Modelling the power spectrum}
\label{sec:modelling_power_spectrum}
In our first model, we describe the impact of varying depth on the power spectrum, following the simplifications described above.
%
%
In accordance with the definition of the shear power spectrum \begin{equation}
\la \hat{\gamma}(\b\ell)\,\hat{\gamma}^*(\b\ell')\ra = (2\pi)^2\delta(\b\ell-\b\ell')P(|\b\ell|) \, ,
\label{eq:original_power_spectrum}
\end{equation}
where $\hat{\gamma}$ denotes the Fourier transform of $\gamma$, we define the observed power spectrum via \begin{equation}
P^{\text{obs}}(\b\ell) \equiv \frac{1}{(2\pi)^2}\int \d^2 \ell'\,  \la \gammaoh(\b \ell)\, \gammaoh {}^*(\b \ell')\ra \, .
\end{equation}
We note that due to the depth-function, both the assumptions of homogeneity and isotropy break down, which means that we can neither assume isotropy in the power spectrum, nor can we assume that $\la \gammaoh(\b \ell) \gammaoh {}^*(\b \ell')\ra$ vanishes for $\b\ell\neq\b\ell'$. This estimator provides a natural extension to the definition of the regular power spectrum and, in the case of a homogeneous depth distribution, is reduced back to the original estimator.
To model a constant depth on each individual pointing, $\b \alpha$, we can choose random variables, $w_{\b \alpha}$, that only need to satisfy $\langle w_{\b \alpha}\rangle=0$. As we assume an infinite number of pointings, $\b\alpha$ can assume any two-dimensional integer value $\mathbb{Z}^2$ and we can parametrize $w(\b\theta)$ as 
\begin{equation}
w(\b \theta) = \sum_{\b \alpha \in \mathbb{Z}^2} w_{\b \alpha} \Xi(\b \theta-L\b \alpha)\, ,
\end{equation}
with the box-function
\begin{equation}
\Xi(\b \theta) = \begin{cases}
1 \qquad \b \theta\in \left[-\frac{L}{2},\frac{L}{2}\right]^2 \\
0 \qquad \text{else}
\end{cases},
\label{eq:defweightf}
\end{equation}
where $L$ is the sidelength of one pointing.
Following the calculations in App.~\ref{sec:calc of PS}, we derive 
\begin{equation}
P^{\text{obs}}(\b \ell)  =  P(\b \ell) + \la w^2\ra \int\frac{\text{d}^2\b \ell'}{(2\pi)^2}\,\hat{\Xi}(\b \ell-\b \ell')\, P(\b \ell')\, .
\end{equation}
Here we have denoted $\la w^2\ra \equiv \la w_{\b \alpha}^2\ra$ as the dispersion of the depth-function, since the statistical properties of this function do not depend on the pointing $\b\alpha$. The Fourier transform of the box function, $\hat{\Xi}$, is a 2-dimensional sinc-function (see \ref{sec:calc of PS}).
The observed power spectrum, $P^{\text{obs}}$, is thus composed of the original power spectrum $P(\ell)$ from Eq. \eqref{eq:original_power_spectrum}, plus a convolution of the power spectrum with a sinc-function, scaling with the variance of the function $w(\b\theta)$. 


\subsection{Modeling the shear correlation functions}
\label{sec:xipm_analytic}
Measures that are more convenient for the inference of cosmological information from observational data are the shear correlation functions $\xi_\pm$, which are defined as \begin{equation}
\xi_\pm(\theta) = \la \gamma_{\rm t}\gamma_{\rm t}\ra(\theta) \pm \la \gamma_\times\gamma_\times\ra(\theta) \, .
\end{equation}
Here, $\gamma_{\rm{t}}$ and $\gamma_\times$, denote the tangential and cross-component of the shear for a galaxy pair with respect to their relative orientation \citep[see][]{2002A&A...396....1S}.
The shear correlation functions are the prime estimators to quantify a cosmic-shear signal, since it is simple to include a weighting of the shear measurements into the correlation functions and, contrary to the power spectrum, one does not have to worry about the shape of the survey footprint or masked regions, or model the noise contribution. 
For this analysis, we follow the assumption that a deeper pointing shows a stronger shear signal $\gamma^{\rm{obs}}(\b\theta) = W(\b\theta)\gamma(\b\theta)$ as described above. This assumption implies that a higher redshift just increases the amplitude of the shear signal, but as can clearly be seen by inspecting shear correlation functions of different redshift distributions, the change of the signal is extremely scale-dependent and not just a multiplication with a constant factor. In other words, not just the average shear changes as a function of redshift, but also its entire two-point statistics. However, this should serve as a reasonable first approximation for small variations in mean source redshift. Additionally, we assume that a greater depth does not only lead to a stronger average shear, but also to a higher galaxy number density, implying a correlation between those two quantities.

\begin{figure}
 \centering
 \includegraphics[width=0.9\linewidth]{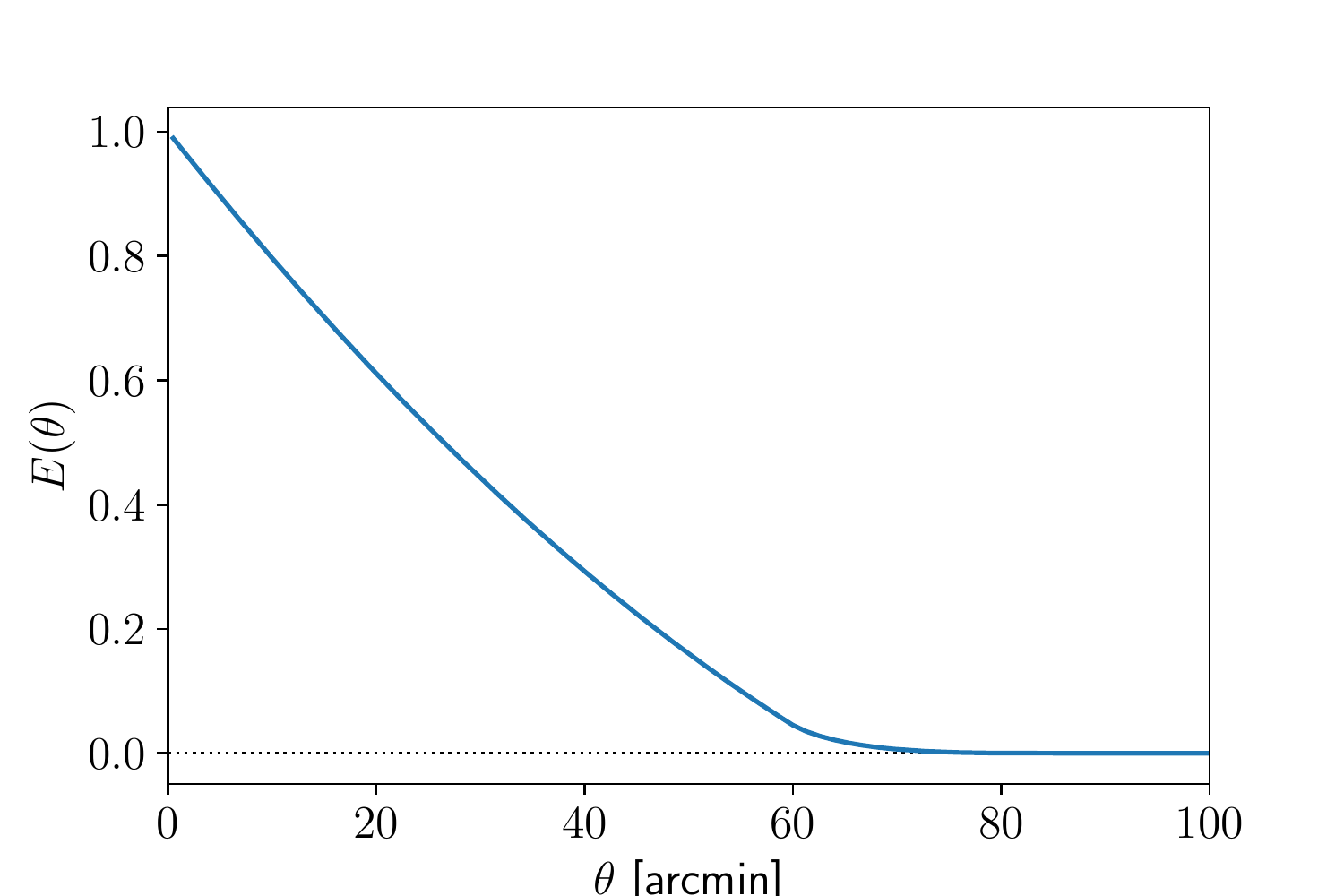}
 \caption{Probability $E(\theta)$ that a random pair of galaxies of distance $\theta$ lie in the same $1\deg ^2$ pointing.}
 \label{fig:eoftheta_lin}
\end{figure}

We denote by $N^i(\b \theta)$ the average weighted number of galaxies\footnote{Instead of using the actual number of galaxies, we take the effective number density, as defined in \citet{Kuijken:2015}, scaled by the respective survey area. Due to this, we account for different weighting of galaxies in the shear correlation functions as well as in the average redshift distribution.} per pointing in redshift bin $i$ and by $W^i(\b \theta)$ the weighting of average shear. The observed correlation functions $\xi^{ij,\text{obs}}_\pm(\theta)$ now change from uniform depth, $\xi_\pm^{ij,\rm{uni}}(\theta)$, via 
\begin{align}
\xi^{ij,\text{obs}}_\pm(\theta) = & \frac{\la N^i(\bth)N^j(\bthp)\gamma^{i,\rm{obs}}_{\rm t}(\bth)\gamma^{j,\rm{obs}}_{\rm t}(\bthp)\ra }{\la N^i(\bth)N^j(\bthp)\ra} \\
 & \pm \frac{\la N^i(\bth)N^j(\bthp)\gamma^{i,\rm{obs}}_\times(\bth)\gamma^{j,\rm{obs}}_\times(\bthp)\ra }{\la N^i(\bth)N^j(\bthp)\ra} \nonumber\\
 = & \frac{\la N^i(\bth)N^j(\bthp)W^i(\bth)W^j(\bthp)\ra}{\la N^i(\bth)N^j(\bthp)\ra} \xi_{\pm}^{ij,\rm{uni}}(\theta) \, ,
 \label{eq:xipmblub1}
 \end{align}
 where the average, $\la\cdot\ra$, represents both an ensemble average as well as an average over the position $\bth$.
 Assuming that the depth of different pointings is uncorrelated, the only important property of a galaxy pair is whether or not they lie in the same pointing. We denote the probability that a random galaxy pair of separation $\theta$ lies in the same pointing by $E(\theta)$. This function is depicted in Fig. \ref{fig:eoftheta_lin}, and an analytic expression is derived in App. \ref{sec:model_e}.
 
To compute the modified shear correlation functions, we parametrize the number densities \linebreak$N^{i}(\b \theta)=\la N^{i} \ra [1+n^{i}(\b \theta)]$ and the weight $W^{i}(\b \theta)=1+w^{i}(\b \theta)$ and, as in Eq. \eqref{eq:defweightf}, interpret $n^{i}(\b \theta)$ as a function with average $\la n^{i} \ra = 0$ that is constant on each pointing. We can see that $\la n^i(\bth)n^j(\bthp)\ra = E(\theta)\la n^i(\bth)n^j(\bth)\ra = E(\theta)\la n^i n^j \ra$ holds and compute:
\begin{align}
&\frac{\la N^i(\bth)N^j(\bthp)W^i(\bth)W^j(\bthp)\ra}{\la N^i\ra \la N^j\ra } \nonumber\\
&\qquad =  1 + \la n^iw^i\ra + \la n^j w^j\ra + E(\theta)\left[ \la n^in^j\ra + \la n^i w^j \ra  + \la n^jw^i\ra \right. \nonumber\\
& \qquad\quad + \la w^iw^j\ra + \la n^in^jw^i\ra + \la n^in^jw^j\ra + \la n^iw^iw^j\ra 
+ \la n^jw^iw^j\ra \nonumber\\
& \qquad\quad\left.+ \la n^in^jw^iw^j\ra  \right] \, .
 \end{align}
Ignoring correlations higher than second order in $n^i$ and $w^i$,\footnote{It is not inherently obvious that this is a valid assumption. However, after performing calculations with and without the inclusion of higher-order correlations, the largest relative difference between the outcomes of both equations was less than $5\times 10^{-4}$.} and performing the same calculation for the denominator of Eq.\,\eqref{eq:xipmblub1}, we find
 \begin{align}
 \xi^{ij,\rm{obs}}_\pm(\theta) = & \left[ 1 + \la n^iw^i\ra + \la n^jw^j\ra + E(\theta)\left(\la n^in^j\ra + \la n^iw^j\ra \right.\right. \nonumber\\
 &\left.\left. + \la n^j w^i\ra + \la w^iw^j\ra\right)\right]\, \left[1+E(\theta)\la n^i n^j\ra \right]\inv\xi^{ij,\rm{uni}}_\pm (\theta) \, .
 \end{align}
 A model correlation function for a cosmic shear survey is usually calculated by taking the average redshift distribution of a redshift bin, weighted by the number density.
 Ignoring that the depth is correlated on scales of one pointing (here at $\theta \leq \sqrt{2}^\circ$) is equivalent to setting $E(\theta)\equiv 0$. We note that there is still a correlation between $N$ and $W$ for the same galaxy.
 Performing the same calculations as above, this yields a relation between the correlation function of uniform depth, $\xi_\pm^{ij,\rm{uni}}$, and the one that is usually modeled, $\xi_\pm^{ij}$:
\begin{equation}
\xi_\pm^{ij}(\theta) = \left(1+\la n^iw^i\ra + \la n^jw^j\ra \right)\xi_\pm^{ij,\rm{uni}}(\theta)\, .
\end{equation}
When an observer now calculates the model correlation functions $\xi^{ij}_\pm$ without accounting for varying depth between pointings, the ratio between modeled and observed correlation functions becomes: \begin{align}
\frac{\xi^{ij}_\pm(\theta)}{\xi_\pm^{ij,\rm{obs}}(\theta)} \approx &  \left[1+\la n^iw^i\ra +\la n^jw^j\ra + E(\theta)\la n^in^j\ra\right] \nonumber\\
& \times \left[1 + \la n^iw^i\ra + \la n^jw^j\ra + E(\theta)\left(\la n^in^j\ra + \la n^iw^j\ra \right.\right. \nonumber\\
& \left.\left. + \la n^j w^i\ra + \la w^iw^j\ra\right)\right]\inv
\, .
\label{eq:xipm_analytic}
\end{align}
It is interesting to note that $\xi^{ij}_\pm = \xi_\pm^{ij,\rm{obs}}$ holds wherever $E(\theta)=0$, so we expect the observed and the modeled correlation functions to be equivalent on scales where the depth is uncorrelated. One thing left to determine is how to define the weight-function $W(\b\theta)$. For this, we will refer the reader to the beginning of Sec.~\ref{sec:results}.


\section{A semi-analytic model}
\label{sec:xipm_semianalytic}
The previously derived analytic model describes, how varying depth between pointings modifies the correlation function due to the correlation between number density and the average redshift of source galaxies. While this model serves as an intuitive first approximation, it completely ignores any effects from the large scale structure (LSS) between the closest and the most distant galaxy. Therefore, we do not expect this model to yield accurate, quantitative results for cosmic shear surveys.

Below we derive a more sophisticated model that includes the effects of the LSS. While it is computationally more expensive, it improves the accuracy of the model for cosmic shear surveys, which are sensitive to the exact redshift distributions of sources as well as the underlying cosmology.

\begin{figure}
\centering
\includegraphics[width=\linewidth]{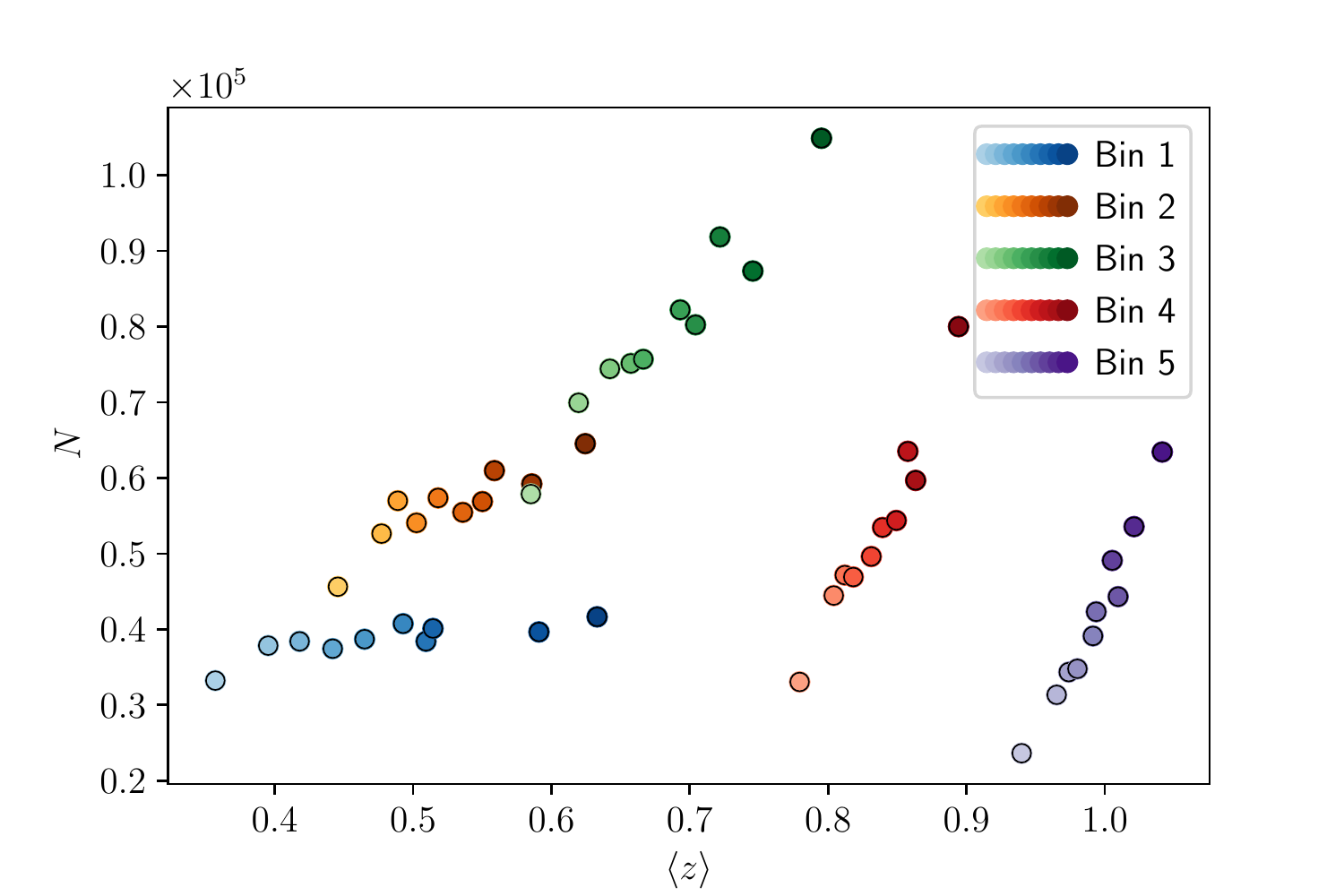}
\caption{Weighted number of galaxies $N$ and average redshift $\la z\ra$ in the KiDS+VIKING-450 survey \citep[KV450,][]{Wright:2018} in pointings of different depth for each of the five tomographic bins used in H18. Each color corresponds to one redshift bin of H18. A single point represents one quantile of the respective redshift bin, where the fainter points denote pointings of shallower depth.}
\label{fig:nz_of_meanz}
\end{figure}

\begin{figure}
\centering
\includegraphics[trim = {1cm 0 1.2cm 0.1cm}, clip, width=\linewidth]{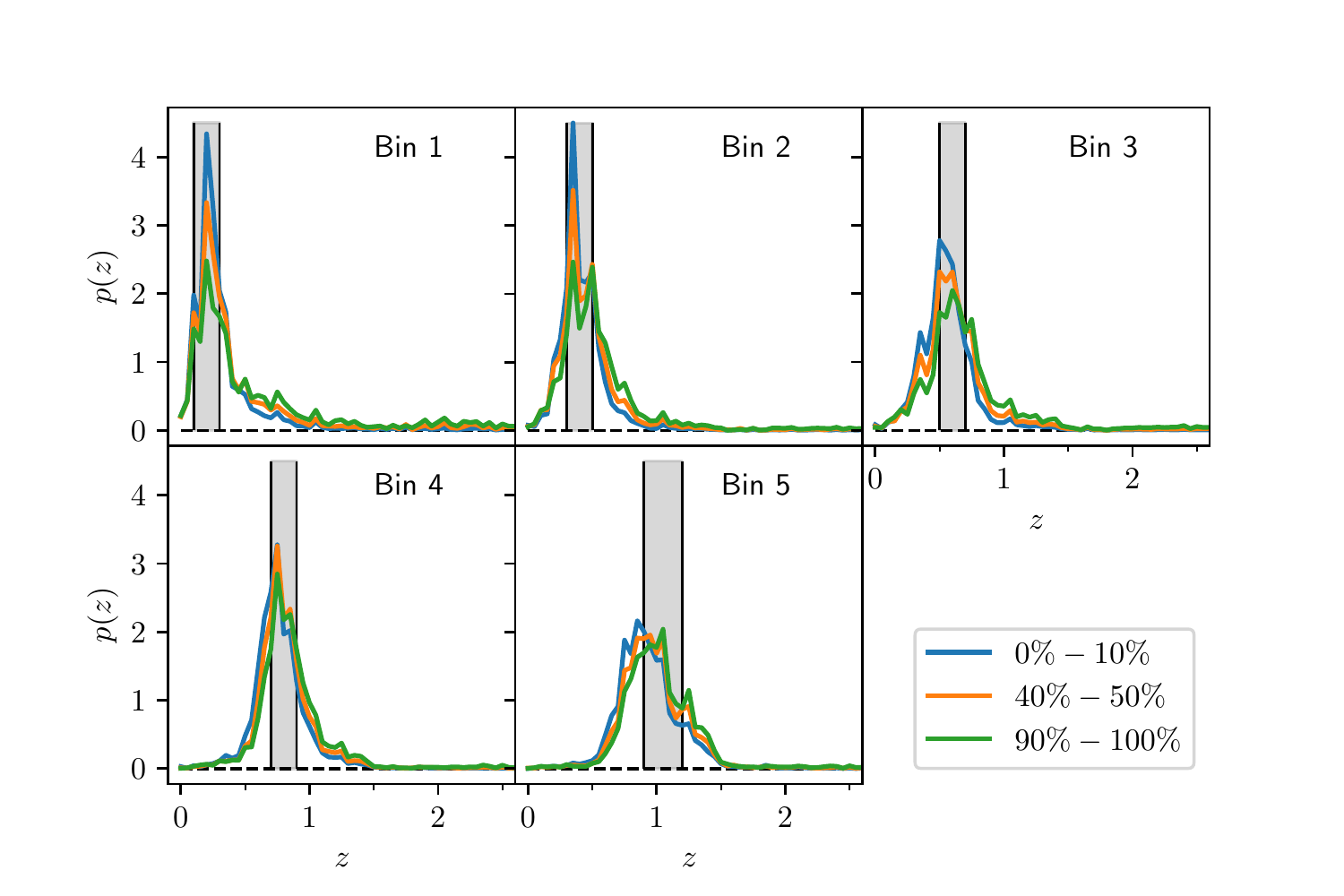}
\caption{Source redshift distributions $p^i_m(z)$ for a selection of very shallow pointings (blue), average pointings (yellow) and very deep pointings (green). The percentage points in the legend denote to which quantile a pointing belongs, when all are ordered by their depth.}
\label{fig:nofz_quantiles}
\end{figure}

An inspection of KiDS-data showed that the redshift distribution of sources is highly correlated with the limiting magnitude in the $r$-band. We thus chose to separate the survey into ten quantiles, sorted by $r$-band depth, that is, if a pointing had a shallower depth than 90\% of the other pointings, it would belong to the first quantile, and so on. For each quantile $m$ and each tomographic redshift bin $i$ we can extract a weighted number of galaxies $N^i_m$ and a source redshift distribution $p^i_m(z)$ following the direct spectroscopic calibration method of H18. In Fig.~\ref{fig:nz_of_meanz}, the average redshift and weighted number of galaxies are plotted for each quantile of each redshift bin, whereas a selection of source redshift distributions is depicted in Fig.~\ref{fig:nofz_quantiles}. A table of the limiting magnitudes for each quantile can be found in Tab.~\ref{tab:maglim}.

Given the two comoving distance probability distributions of sources, $\mathcal{L}^i_m(\chi)$ and $\mathcal{L}^j_n(\chi)$, we can compute the shear correlation functions from the underlying matter power spectrum, $P_\delta(k,\chi)$, via \citep{1992ApJ...388..272K} \begin{align}
\label{eq:xipm-pkappa}
\xi_{\pm,mn}^{ij}(\theta) =& \int_0^\infty \frac{{\rm d}\ell\,\ell}{2\pi}\, J_{0,4}(\ell\theta)\, P^{ij}_{\kappa,mn}(\ell)\, , \\
\label{eq:pkappa-pdelta/lenseff}
P^{ij}_{\kappa,mn}(\ell) =& \frac{9 H_0^4\Omega_{\rm m}^2}{4c^4}\int_0^{\chi_{\rm{H}}} {\rm d}\chi\, \frac{g^i_m(\chi)g^j_n(\chi)}{a^2(\chi)}\, P_\delta\left(\frac{\ell}{f_K(\chi)},\chi\right)\, , \\
\label{eq:lenseff}
g^i_m(\chi) =& \int_\chi^{\chi_{\rm{H}}} {\rm d}\chi' \, \mathcal{L}^i_m(\chi') \, \frac{f_K(\chi'-\chi)}{f_K(\chi')}\, .
\end{align}
Here, $J_n$ denotes the $n$-th order Bessel Functions, $f_K(\chi)$ is the comoving angular diameter distance and $\chi_{\rm{H}}$ is the comoving distance to the horizon. The parameters $H_0$ and $c$ denote the Hubble constant and the speed of light.

Using Eq.~\eqref{eq:xipm-pkappa}, we can compute the model correlation functions, $\xi_{\pm,mn}^{ij}(\theta)$, for each pair of quantiles $m,n$ and redshift bins $i,j$.\footnote{For the calculation of the shear correlation functions we use \textsc{Nicaea} \citep{2009A&A...497..677K}. 
For the power spectrum on nonlinear scales, we use the method of \citet{2012ApJ...761..152T}.} When measuring the shear correlation functions of a survey, we take the weighted average of tangential and cross shears of all pairs of galaxies \citep[see][]{2017MNRAS.465.1454H}. If, for a single pair of galaxies, one galaxy lies in the $m$-th quantile of redshift bin $i$ and the second one lies in the $n$-th quantile of redshift bin $j$, then their contribution to the observed correlation functions is, on average, $\xi_{\pm,mn}^{ij}(\theta)$. This means that, if we know each of those single correlation functions, we can reconstruct the total correlation functions via a weighted average of the single functions. Formally, we define \begin{equation}
\xi_\pm^{ij,\rm{obs}}(\theta) = \frac{\sum_{m,n} \mathcal{P}_{mn}^{ij}(\theta)\,\xi_{\pm,mn}^{ij}(\theta)}{\sum_{m,n} \mathcal{P}_{mn}^{ij}(\theta)}\, ,
\label{eq:def_xiobs}
\end{equation}
where $\mathcal{P}_{mn}^{ij}$ is a weighting of the correlation functions, which has to be proportional to the probability that a galaxy pair of separation $\theta$ comes from quantiles $m$ and $n$. In this analysis, we will assume an uncorrelated distribution of depth and neglect boundary effects as well as the sample variance of the depth-distribution between pointings. We will later discuss the validity of these assumptions as well as possible mitigation strategies. 

To calculate $\mathcal{P}_{mn}^{ij}(\theta)$, we imagine two arbitrary (infinitesimally small) surface elements $\d^2\b\theta_1$ and $\d^2\b\theta_2$ of separation $\theta$ on the sky. For the case $m\neq n$, we know that the two galaxies contributing to $\mathcal{P}_{mn}^{ij}(\theta)$ have to lie in different pointings, else they would automatically be in the same quantile. The probability that the surface elements are within different pointings is $[1-E(\theta)]$. Furthermore, the first element $\d^2\b\theta_1$ has to lie in quantile $m$, the probability of which is $1/10$. The pointing of the second element $\d^2\b\theta_2$ has to be of quantile $n$; the probability of that is also equal to $1/10$. The probability that a galaxy pair populates those surface elements is proportional to the weighted number of galaxies $N_m^i$ and $N_n^j$. We get for $n\neq m$: \begin{equation}
\mathcal{P}_{mn}^{ij}(\theta) = [1-E(\theta)]\frac{1}{100} N_m^i N_n^j\, .
\label{eq:pmnij_corr1}
\end{equation}
For the calculation of $\mathcal{P}_{mm}^{ij}(\theta)$, we have to account for a different possibility: In case that the galaxies lie in the same pointing, they automatically are in the same quantile. We therefore obtain \begin{equation}
\mathcal{P}_{mn}^{ij}(\theta) = E(\theta)\frac{1}{10} N_m^iN_m^j\,\delta_{mn} + [1-E(\theta)]\frac{1}{100} N_m^i N_n^j \, ,
\label{eq:pmnij_uncorr}
\end{equation}
where $\delta_{mn}$ denotes the Kronecker delta.
Inserting this into Eq.\,\eqref{eq:def_xiobs}, we compute
\begin{align}
\xi_{\pm,mn}^{ij,\rm{obs}}(\theta) =  \left.\frac{1}{C}\sum_{m=1}^{10} N_m^i \right\{ & E(\theta) N_m^j \xi_{\pm,mm}^{ij}(\theta) \nonumber\\
 & \left. + \frac{\big[1-E(\theta)\big]}{10}\sum_{n=1}^{10}N_n^j \xi_{\pm,mn}^{ij}(\theta)\right\}\, ,
\label{eq:correctionfunction1}
\end{align}
with the normalization
\begin{equation}
C = \sum_{m=1}^{10} N_m^i \left[ E(\theta)  N_m^j + \frac{\big[1-E(\theta)\big]}{10}\sum_{n=1}^{10} N_n^j\right]\, .
\end{equation}
A mathematically more rigorous derivation of this function can be found in App.~\ref{sec:calc of xipm}.

Computing this for all five redshift bins of the KV450-survey, forces us to calculate and co-add 1275 correlation functions\footnote{For each of the 15 pairs of redshift bins we need to calculate 100 correlation functions, except for the pairs of bins with the same redshift, where only 55 correlation functions need to be calculated due to symmetry.}. Since the variation in depth is a relatively small effect, even tiny numerical errors can add up, skewing the calculations. Additionally, calculating $10^3$ correlation functions is computationally expensive. However, if we examine Eq.~\eqref{eq:lenseff}, we see that the comoving distance distribution of sources enters linearly. This, in turn, implies that in Eqs.~\eqref{eq:pkappa-pdelta/lenseff} and \eqref{eq:xipm-pkappa}, both source distance distributions enter linearly, meaning that, instead of adding correlation functions, we can add their respective redshift distributions and compute the correlation functions of that. In particular, we can define the \textit{combined number of galaxies} $N^i$ and \textit{average comoving distance probability distribution}, $\mathcal{L}^i(\chi)$, of tomographic bin $i$ as \begin{equation}
N^i\equiv\sum_m N_m^i\, , \qquad \mathcal{L}^i(\chi) = \frac{\sum_m N_m^i \mathcal{L}_m^i(\chi)}{\sum_m N_m^i} \, .
\end{equation}
Defining $\xi^{ij}_\pm$ as the correlation functions between the average comoving distance distributions $\mathcal{L}^i(\chi)$ and $\mathcal{L}^j(\chi)$, we find: \begin{align}
& \sum_{m,n}N_m^iN_n^j\xi^{ij}_{\pm,mn}(\theta) = \, \frac{9 H_0^4\Omega_{\rm m}^2}{4c^4} \sum_{m,n} N_m^i N_n^j \int_0^\infty \frac{{\rm d}\ell\,\ell}{2\pi}\, J_{0,4}(\ell\theta) \nonumber\\
& \qquad\times \int_0^{\chi_{\rm{H}}} \frac{{\rm d}\chi}{a^2(\chi)}\, P_\delta\left(\frac{\ell}{f_K(\chi)},\chi\right)
\int_\chi^{\chi_{\rm{H}}} {\rm d}\chi' \, \mathcal{L}^i_m(\chi') \, \frac{f_K(\chi'-\chi)}{f_K(\chi')} \nonumber\\
& \qquad\times \int_\chi^{\chi_{\rm{H}}} {\rm d}\chi'' \, \mathcal{L}^j_n(\chi'') \, \frac{f_K(\chi''-\chi)}{f_K(\chi'')} \nonumber\\
 & \quad =  \,N^i N^j \, \frac{9 H_0^4\Omega_{\rm m}^2}{4c^4} \int_0^\infty \frac{{\rm d}\ell\,\ell}{2\pi}\, J_{0,4}(\ell\theta) \int_0^{\chi_{\rm{H}}} \frac{{\rm d}\chi}{a^2(\chi)}\, P_\delta\left(\frac{\ell}{f_K(\chi)},\chi\right) \nonumber\\
& \qquad\times \int_\chi^{\chi_{\rm{H}}} {\rm d}\chi' \, \frac{\sum_m N_m^i \mathcal{L}^i_m(\chi')}{N^i} \, \frac{f_K(\chi'-\chi)}{f_K(\chi')} \nonumber\\
& \qquad\times \int_\chi^{\chi_{\rm{H}}} {\rm d}\chi'' \, \frac{\sum_n N_n^j \mathcal{L}^j_n(\chi'')}{N^j}  \, \frac{f_K(\chi''-\chi)}{f_K(\chi'')} \nonumber\\
& \quad =  \, N^iN^j\xi^{ij}_\pm(\theta)\, .
\end{align}
Consequently, we can apply this to Eq.~\eqref{eq:correctionfunction1}, yielding
\begin{align}
\xi_{\pm}^{ij,\rm{obs}}(\theta) = & \frac{1}{C}\left\{ E(\theta)\left[\sum_{m=1}^{10} N_m^iN_m^j \xi_{\pm,mm}^{ij}(\theta)\right]\right.\nonumber\\
& \left.\vphantom{\sum_{m=1}^10}+\frac{\big[1-E(\theta)\big]}{10}\xi_\pm^{ij}(\theta)N^iN^j\right\}\, .
\label{eq:correctionfunction2}
\end{align}
For each pair of redshift bins we, thus, only have to compute eleven correlation functions, which reduces the number of functions to compute from 1275 to 165.


\section{Results}
\label{sec:results}
We compare the analytic and semi-analytic models for a variable-depth cosmic shear measurement in a KiDS-like survey. While the application of the semi-analytic method is straightforward, for the analytic method we need to decide how to estimate the weight function $W$ from the given redshift data. Following the separation of a survey into quantiles as in Sect. \ref{sec:xipm_semianalytic}, we define $W(\b\theta)\equiv W_n$ whenever $\b\theta$ is in a pointing of quantile $n$. For the determination of $W_n$ we test two approaches:
 As a first method, following \citet{2006APh....26...91V,1997A&A...322....1B}, we estimate 
\begin{equation}
W_n \propto \la z \ra _n^{0.85}\, ,
\end{equation}
where $\la z\ra_n$ is the average redshift of quantile $n$. As a second method, we define \begin{equation}
W_n \propto \sqrt{ \la\gamma_{\rm{t}}\gamma_{\rm{t}}\ra (\theta_{\rm{ref}}) + \la \gamma_\times\gamma_\times\ra (\theta_{\rm{ref}})} = \sqrt{\xi_{+,nn}^{ij}(\theta_{\rm{ref}})}\, ,
\end{equation}
where the $\xi_{+,nn}^{ij}(\theta_{\rm{ref}})$ denotes the model correlation function defined in Sect. \ref{sec:xipm_semianalytic}, evaluated at a characteristic scale $\theta_{\rm{ref}}$, that needs to be chosen.

While the first method suffers from the fact that the power-law index only holds for sources of redshifts $1\lesssim z \lesssim 2$, the second method is sensitive to the angular range $\theta_{\rm{ref}}$, at which the shear correlation functions are evaluated, which is fairly arbitrary. For $\theta_{\rm{ref}}\approx 11'$, which is roughly in the logarithmic middle between the range of the correlation functions, $[0.\!'5,300']$, the two calibration methods agree. The choice of other values for $\theta_{\rm{ref}}$ leads to a different amplitude of the change $\xi_\pm/\xi_\pm^{\rm{obs}}$, but does not affect its shape. Generally, a smaller $\theta_{\rm{ref}}$ leads to a stronger effect, in particular, the highest amplitude of the change is at $\theta_{\rm{ref}}=0.\!'5$.

\subsection{Effect on the shear correlation functions}

\begin{figure*}
  \centering
  \includegraphics[width=0.9\linewidth]{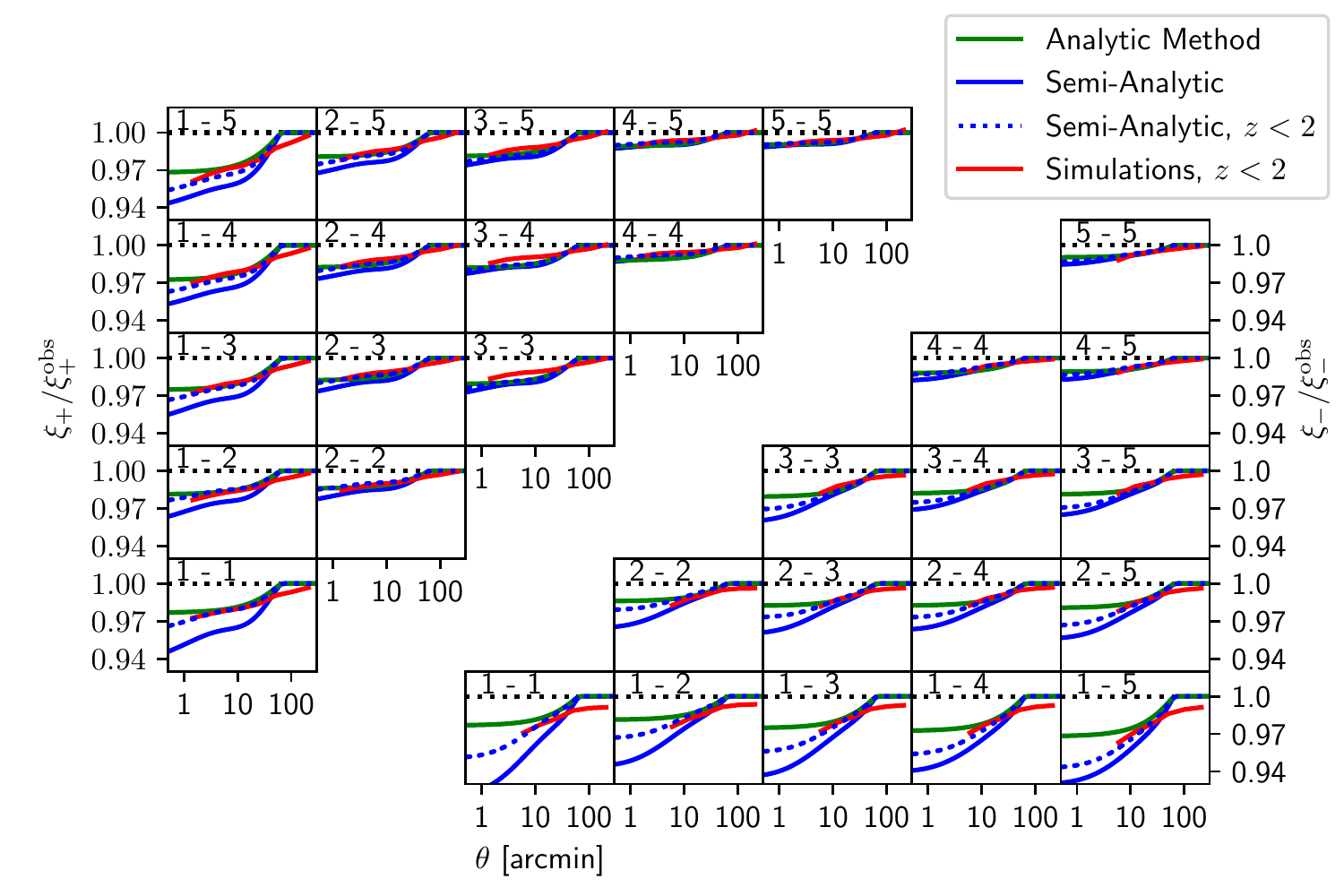}
  \caption{Ratio of correlation functions measured for a uniform depth survey, $\xi_\pm$, and a KiDS-like variable depth survey, $\xi_\pm^{\rm{obs}}$, cross-correlating five tomographic bins (as denoted in the upper left corner of each panel). The upper left triangle depicts the ratios of $\xi_+$, whereas the lower right triangle depicts the ratios of $\xi_-$. Results from mock KiDS-like data (red) can be compared to analytic models from Sect.~2 (average redshift weighting, green), and the semi-analytic model from Sect.~3 (blue solid). Mock data is limited to the angular regime which is not significantly impacted by resolution effects. As the mocks only take galaxies with $z<2$ into account, they slightly underestimate the effect. Applying the same redshift-cutoff to the semi-analytical model (blue dashed) yields a near-perfect agreement on sub-pointing scales. Therefore, the seemingly better agreement between the mocks and the analytic method is purely coincidental. The models adopt the best-fit cosmology of H18.}
  \label{fig:all_xis}
  \end{figure*}

In this Section, we calculate both the analytic (Eq.~\ref{eq:xipm_analytic}) and semi-analytic (Eq.~\ref{eq:correctionfunction2}) models for the shear correlation function from a KiDS-like variable depth survey.   We adopt the tomographic bins defined in H18 and their resulting best-fit cosmological parameters to present, in Fig.~\ref{fig:all_xis}, the ratio between our models for the observed correlation functions $\xi_\pm^{\rm{obs}}$, and the standard theoretical prediction that assumes uniform depth. We find that the level of variation in the depth of the KiDS survey increases the amplitude of the observed shear correlation function, on sub-pointing scales, by up to 5\% relative to the uniform-depth case.

We compare our models to mock KiDS-like data created using a modified version of the Full-sky Lognormal Astro-fields Simulation Kit \citep[\textsc{FLASK},][Joachimi, Lin, et al., in prep.]{Xavier:2016}.  Using {\sc{FLASK}} lognormal fields, we generate galaxy mocks with coherent clustering and lensing signals.   Adopting a linear relation between the limiting $r$-band magnitude and the effective number density, fit to the KiDS data, we imprint the variable depth of the full KiDS-1000 footprint \citep{Kuijken:2019} on a {\sc{Healpix}}\footnote{\url{https://healpix.sourceforge.io}} grid of $\text{Nside}=4096$.  Our resolution choice represents a compromise between minimising the mock computation time and maximising the accuracy of the recovered shear signal at small angular scales. With $\text{Nside}=4096$, $\xi_+$ is accurate to 7\% above $\sim 1$ arcmin, and $\xi_-$ is accurate to 10\% above $\sim6$ arcmin.   For each {\sc{Healpix}} pixel in the KiDS-1000 data, the limiting $r$-band magnitude defines the effective number density of sources, and the average source redshift distribution for each tomographic redshift bin (see Fig.~\ref{fig:nofz_quantiles}).  In the uniform depth case, redshifts and number densities are sampled from the average of these tomographic sets.  The ratio of the lensing signals from the two mocks is computed, averaged over 2000 shape-noise-free realizations, and is shown in Fig.~\ref{fig:all_xis} (red).

The results can be seen in Fig.~\ref{fig:all_xis}.
We observe that for high-redshift bins, all methods yield consistent results. For low-redshift bins, there are discrepancies between the different models. However, the average-redshift weighted analytic model, as shown here, is only valid for high redshifts, whereas the auto-correlation-$\xi_+$-weighted model (not shown) is entirely dependent on the choice of $\theta_{\rm{ref}}$. For $\theta_{\rm{ref}}=11'$, both analytic models agree very well for all redshift bins. For $\theta_{\rm{ref}}=0.\!'5$, the auto-correlation weighted method agrees with the semi-analytic one pretty well for $\xi_+$. As $\xi_-$ is affected much stronger by this effect\footnote{The effect on $\xi_-$ is much stronger due to the fact that in Equation \eqref{eq:xipm-pkappa}, $\xi_+$ is computed by filtering the power spectrum with the 0-th order Bessel function. This function peaks at $\ell\theta=0$, meaning that for all values of $\theta$, the correlation function $\xi_+$ is sensitive to small values of $\ell$, corresponding to large separations $\theta$. However, $\xi_-$ is obtained by filtering with the 4-th order Bessel function, which peaks at approximately $\ell\theta\approx 5$, so for different $\theta$ this function is sensitive to varying parts of the convergence power spectrum. A more detailed analysis of this can be found in the Appendix of \citet{2017MNRAS.471.4412K}.}, the analytic method is not able to trace this change for any choice of $\theta_{\rm{ref}}$. Furthermore we note, that the effect seems to be strongest in the first redshift bin, which is not surprising, as there the average redshift between pointings varies the most (compare Fig. \ref{fig:nz_of_meanz}).

The simulations and the models seem to be in relatively good agreement, but there are some differences. It is noticeable that in the simulations, the value $\xi_\pm/\xi_\pm^{\rm{obs}}$ consistently stays below unity at large scales, which can be attributed to the fact that the depth of different pointings is not completely uncorrelated, as was assumed in the models. 

An additional difference between the models and simulations is, that in the models we neglect boundary effects and the sample-variance of the depth-distribution between pointings, whereas the simulations were performed with the KiDS-1000 footprint. In App.~\ref{sec:expand_eoftheta}, we develop a model to extract the correction $\xi_\pm^{ij}/\xi_{\pm}^{ij,\rm{obs}}$ for a specific survey footprint. With this model, we can estimate the impact of a correlated distribution of depth, the sample variance of the depth-distribution and boundary effects. We find that for a square footprint of $450\,\rm{deg}^2$ or $1000\,\rm{deg}^2$ with an uncorrelated depth-distribution, finite field effects are negligible.

In general, the semi-analytic model predicts a stronger effect than the mocks. This is due to the fact that the mocks are subject to a redshift-cutoff at $z=2$, meaning that they do not take the high-redshift tail of galaxies into account. This is of particular importance in the first redshift bin, where this feature is especially pronounced (compare Fig.~\ref{fig:nofz_quantiles}). When the same cutoff is applied to the models, the agreement on sub-pointing scales is striking. In particular this means that the mocks slightly underestimate the effect of varying depth.

\begin{figure*}
$
\begin{array}{cc}
\includegraphics[width=0.45\textwidth]{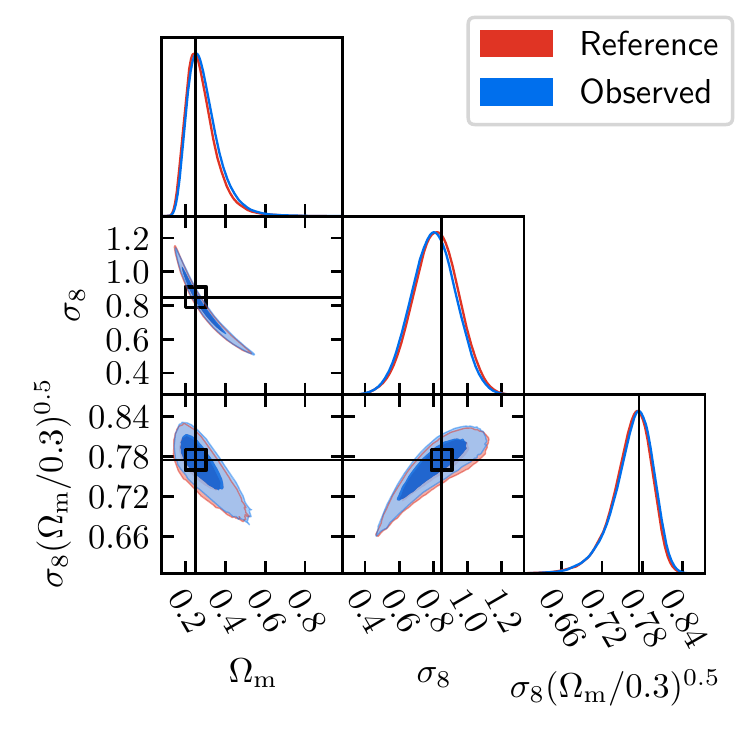}&
\includegraphics[width=0.45\textwidth]{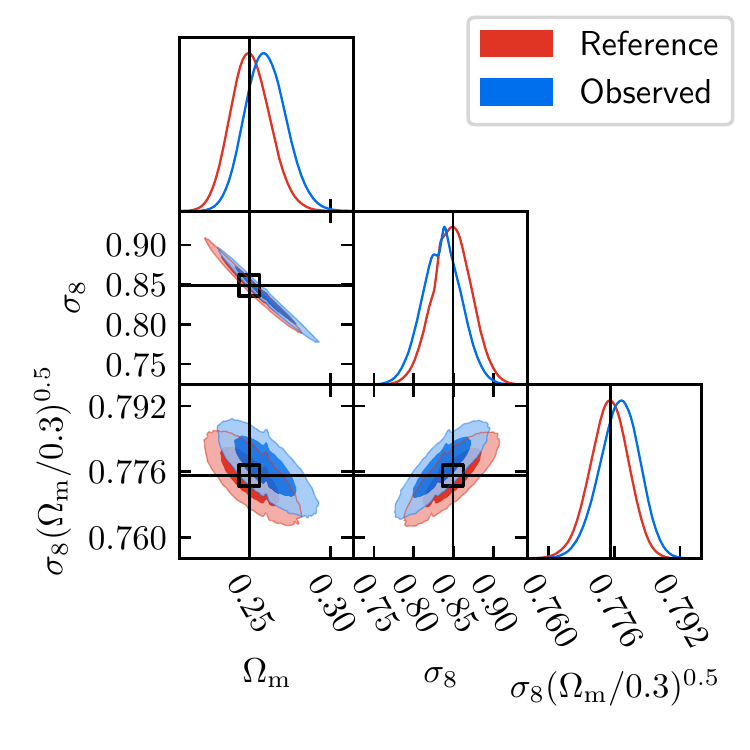}
\end{array}$
\caption{Recovered cosmological parameters for a variable-depth (Observed) and uniform-depth (Reference) KiDS-$450$-like (left) and KiDS-$15\,000$-like survey (right). The KiDS-$450$-like figure was computed using the covariance matrix of H18. For the KiDS-$15\,000$-like survey, we divided the covariance matrix of H18 by 30. This approximately accounts for the increased survey area of next-generation experiments, but does not factor in the increased number density and higher redshifts. Hence, this exercise provides a rough indication of the significance of varying depth effects in stage IV surveys. Both figures were computed using a fiducial cosmology of $\Omega_{\rm{m}}=0.25$ and $\sigma_8 = 0.85$.}
\label{fig:mcmc_results}
\end{figure*} 

\subsection{Impact on cosmological parameter constraints}

As the next step, we assess how the observational depth variations propagate to cosmological parameters inferred from $\xi_\pm^{ij,\rm{obs}}$ compared to $\xi_\pm^{ij}$. For this test, we choose a fiducial cosmology, $\b \Phi$, and determine the relative change in $\Omega_{\rm m}$ and $\sigma_8$ compared to a reference setup with uniform depth. All other cosmological parameters are kept fixed. First, we compute the reference correlation functions, $\xi_\pm^{ij}(\theta,\b\Phi)$, for each pair of redshift bins $i,j$ using \textsc{Nicaea} as described in Sec.~\ref{sec:xipm_semianalytic}. Then we derive the observed correlation functions, $\xi_\pm^{ij,\rm{obs}}(\theta,\b\Phi)$, from Eq.~(\ref{eq:correctionfunction2}). Using the Markov-Chain Monte Carlo sampler \textsc{emcee}, we sample correlation functions $\xi_\pm^{ij}(\theta,\b\Phi')$ for different cosmologies $\b\Phi'$ and find the likelihood distribution, given the data vector $\xi_\pm^{ij,\rm{obs}}(\theta,\b\Phi)$ and the covariance-matrix computed in H18. This yields an estimate of the shift in $\Omega_{\rm m}$ and $\sigma_8$ introduced by varying depth.

As can be seen in Fig.~\ref{fig:mcmc_results}, the impact of varying depth is insignificant compared to the uncertainties for a KiDS-$450$-like survey. To get a rough estimate for the impact on future surveys, we divide our covariance-matrix by a factor of 30 to model a KiDS-$15\,000$-like survey, which approximately accounts for the increased survey area of LSST and Euclid with respect to KiDS-450. Here the impact on $\Omega_{\rm m}$, $\sigma_8$ and $S_8$ is significant at the level of approximately $1\sigma$. As our modified covariance-matrix does not account for the factor of $\sim \! 4$ expected increase in galaxy number density for LSST and Euclid, we note that this is likely to be a lower estimate for the significance of the effect.
 Even though Euclid is a space-based mission and, therefore, will not suffer from variable atmospheric effects, the key photometric redshift measurement uses data from several ground-based surveys, including KiDS.   Placing a selection criteria on redshift estimation success, will therefore lead to depth variations in the source galaxy sample. While the data from LSST will be practically free of variations in depth after 10 years of observations, the first few years data will include significant depth variation. The impact may be even stronger than the KiDS-like analysis presented here as the multi-band KiDS depth variation was minimised using seeing-dependent data acquisition. This is in contrast to the seeing-agnostic multi-band cadence of LSST.

Calculating the correction $\xi_\pm^{ij}/\xi_\pm^{ij,\rm{obs}}$ for varying values of $\Omega_{\rm{m}}$ and $\sigma_8$, reveals a nontrivial dependence on the cosmology, which can be seen in Fig.~\ref{fig:comparecosmo}. For various combinations of $\Omega_{\rm{m}}$ and $\sigma_8$ within the $95\%$ confidence limit of KV450, we report a variation in $\xi_\pm/\xi_\pm^{\rm{obs}}$ of a few percentage points on small scales.

\subsection{Variable depth contribution to B-modes}

\begin{figure*}
\centering
\includegraphics[width = 0.9\linewidth]{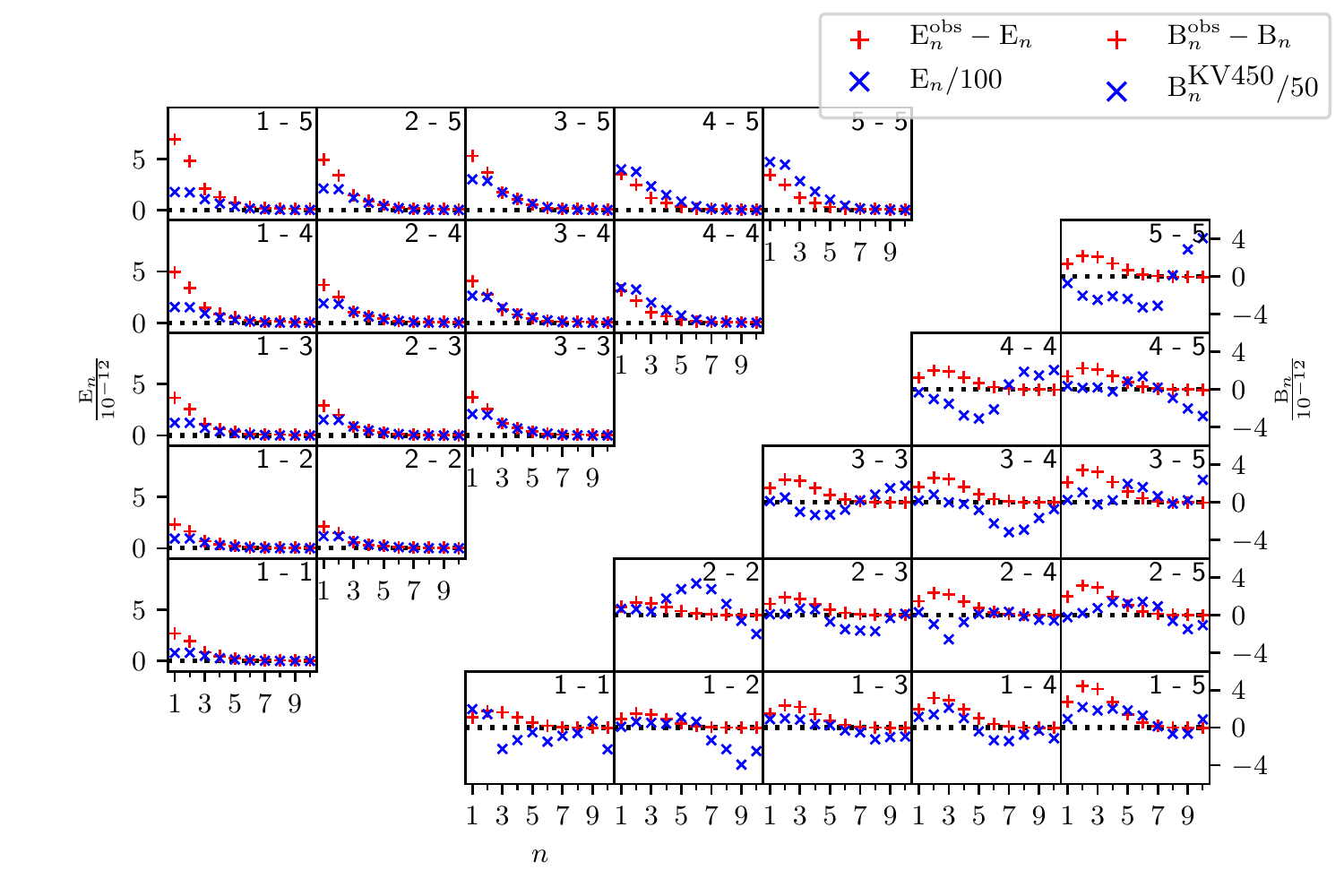}
\caption{Difference in the E-modes (top left) and B-modes (bottom right) between the reference and the observed correlation functions. For comparison: Scaled total E-modes of the reference correlation function $E_n$ and scaled B-modes measured in the KV450 survey $B_n^{\rm{KV450}}$. All E- and B-modes were calculated using the logarithmic COSEBIs in S10 for an angular range of $\theta_{\rm{min}}=0.\!^\prime 5,\,\theta_{\rm{max}}=72\arcmin$.}
\label{fig:bmodes_cosebi}
\end{figure*}

To check for remaining systematics, a weak lensing signal can be divided into two components, the so-called E- and B-modes \citep{2002ApJ...568...20C,2002A&A...389..729S}. To leading order, B-modes cannot be created by astrophysical phenomena and are thus an excellent test for remaining systematics. Direct E- and B-mode decomposition for cosmic shear surveys can be provided by Complete Orthogonal Sets of E- and B-mode Integrals \citep[COSEBIs,][hereafter S10]{2010A&A...520A.116S}, as they can easily be applied to real data. We note that the non-existence of B-modes does not necessarily imply that the sample is free of remaining systematics.
To estimate the B-modes created by this effect, we extract the COSEBIs
 from the correlation functions, $\xi_\pm^{\rm{obs}}$, that have been modified under the semi-analytic model, and from a reference set of correlation functions $\xi_\pm$. To be most sensitive to the effect of varying depth, we choose logarithmic COSEBIs with an angular range of $\theta_{\rm{min}}=0.\!'5$ to $\theta_{\rm{max}}=72'$. As the B-modes of the reference correlation functions are zero, they are a good test for numerical errors in the calculations. Motivated by the discussion in \citet{2017MNRAS.464.1676A}, we calculated the COSEBIs from correlation functions binned in $400\,000$ linear bins, which yielded neglible numerical errors. We report a consistent B-mode behaviour across all redshift-bins, which can be seen in Fig.~\ref{fig:bmodes_cosebi}. However, we compare the B-Modes to the ones measured by \citet{asgari/etal:2019} in KV450, which were consistent with zero. Since the B-modes created by varying depth are smaller than these by a factor of 50, we conclude that this effect cannot create measurable B-modes in the KV450 survey. It should be noted that the difference in E-modes is as large as the B-modes, which suggests that any significant change in the cosmological parameters due to varying depth will also yield a significant detection of B-modes. Additionally, the created pattern is very characteristic, which makes it easy to recognize in a B-mode analysis of an actual survey \citep[see][]{2018arXiv181002353A}.


\section{Discussion}
\label{sec:discussion}
With our semi-analytic model we describe the impact of varying depth in ground-based cosmic shear surveys. During our analysis we made several simplifications, which we discuss below.

In the most general terms, we analyze the effects of a position-dependent selection function on cosmic shear surveys. In our analysis, this selection function was governed by the KiDS $r$-band depth of a pointing. This neglects a number of other effects: The depth in different bands and the seeing of a pointing will also modify the number densities and redshift distributions on the scale of a pointing (although those variations are also correlated with $r$-band depth and thus at least partly accounted for), whereas dithering strategies as well as imperfections in the telescope and CCD cause modifications on sub-pointing scales. However, several tests showed that these effects are subdominant compared to the variations caused by the $r$-band depth.

We assumed an uncorrelated distribution of the depth-function and neglected boundary effects as well as the sample variance of the depth-distribution between pointings. While the boundary effects arising from a finite survey footprint have a small impact on the shape of the function $E(\theta)$\footnote{This would be due to the fact that a pointing next to a boundary has fewer neighbours, therefore making it more likely that a galaxy pair is in the same pointing.}, the governing factor is the sample variance of the depth-distribution. We assumed that the probability for any pointing to be in quantile $n$ is exactly the expectation value, namely $1/10$. While this would be true for an infinitely large survey with an uncorrelated distribution of the depth-function, it does not necessarily hold for a real survey. However, our analysis in App.~\ref{sec:expand_eoftheta} suggests that these effects are not significant for the KV450 survey. In the models, we also assumed an uncorrelated distribution of the depth-function. As can be seen in Fig \ref{fig:all_xis}, this approximation introduces a small error when compared to the simulations.

In our MCMC runs, we did not account for degeneracies with other cosmological parameters or observational effects. In particular, intrinsic alignments and baryon feedback also modify the correlation functions primarily on small scales, so they are likely to be degenerate with the effect of varying depth \citep{Troxel:2015}. In an MCMC run that includes these nuisance parameters, we suspect that the parameters for intrinsic alignments and baryon feedback change to mitigate this effect, so that the impact on cosmological parameters will be smaller than in our results. 

Despite these repercussions, we are confident to say that the effects of varying depth are not significant for the KV450 survey. In particular, this means that a varying depth cannot explain the tension between observations of the low-redshift Universe and results from analysis of the CMB.

For next-generation surveys like Euclid and LSST, we have demonstrated that if variable depth is unaccounted for in the analysis, the resulting bias is likely to be significant. A detailed LSST and Euclid study that uses a realistic variation of depth should therefore be conducted. If these studies reach a similar conclusion, variable-depth bias could be circumvented in likelihood analyses by including a cosmology-dependent correction for this effect using the semi-analytical model presented in this paper (see Fig. \ref{fig:comparecosmo}). 

Although the analytic model (Eq.~\ref{eq:xipm_analytic}) does not fully describe the effect, it can be used as a valid approximation to estimate the importance of varying depth for an arbitrary survey: As can be seen in Fig.~\ref{fig:nz_of_meanz}, in the KiDS-survey the characteristic changes both in redshift and number density for the third redshift bin are about $0.1$. Setting $\la n^in^j\ra=\la w^iw^j\ra=\la n^iw^j\ra = 0.01$ yields $\xi^{ij}_\pm/\xi_\pm^{ij,\rm{obs}}(0)\approx 0.97$, which roughly agrees with the actual results (see Fig.~\ref{fig:all_xis}). For a survey with only half the variation in depth, one gets $\xi^{ij}_\pm/\xi_\pm^{ij,\rm{obs}}(0)\approx 0.993$. This method allows us to estimate a threshold for an acceptable variation of depth, given a required precision for the shear correlation functions.

Additionally, it is interesting to note that $E(\theta)$ is the azimuthal average of the function $E(\b\theta)$ derived in Sect.~\ref{sec:model_e}, which is not isotropic. Therefore, it would be possible to observe a direction-dependent correlation function $\xi_\pm^{ij,\rm{obs}}(\b\theta)$ in future surveys. An anisotropy in the observed correlation function could be a sign for the influence of varying depth.

The variations in depth will also affect the covariance of a survey, both because they modify the signal and because they introduce an additional term of sample variance in terms of the distribution of depth. This effect will be investigated in a forthcoming publication (Joachimi, Lin, et al., in prep.).


\begin{acknowledgements}
We thank the anonymous referee for very constructive and helpful comments.
The results in this paper are based on observations made with ESO Telescopes at the La Silla Paranal Observatory under programme IDs 177.A-3016, 177.A-3017, 177.A-3018 and 179.A-2004, and on data products produced by the KiDS consortium. The KiDS production team acknowledges support from: Deutsche Forschungsgemeinschaft, ERC, NOVA and NWO-M grants; Target; the University of Padova, and the University Federico II (Naples).
We acknowledge support from the European Research Council under grant numbers 770935 (HH,JLvdB) and 647112 (CH,MA,CL).
SH acknowledges support from the German Research Foundation (DFG SCHN 342/13).
H. Hildebrandt is supported by a Heisenberg grant of the Deutsche Forschungsgemeinschaft (Hi 1495/5-1).
CH acknowledges support from the Max Planck Society and the Alexander von Humboldt Foundation in the framework of the Max Planck-Humboldt Research Award endowed by the Federal Ministry of Education and Research. 
KK acknowledges support by the Alexander von Humboldt Foundation.
TT acknowledges funding from the European Union’s Horizon 2020 research and innovation programme under the Marie Sk{l}odowska-Curie grant agreement No 797794.
Some of the results in this paper have been derived using the HEALPix \citep{2005ApJ...622..759G} package.
This research has made use of NASA's Astrophysics Data System and adstex (\url{https://github.com/yymao/adstex}).
\\
\emph{Author contributions:} All authors contributed to the development and writing of this paper. The authorship list is given in two groups: the lead authors (SH, PS, HHi) followed by key contributors to the scientific analysis in alphabetical order.
\end{acknowledgements}

\bibliographystyle{aa}
\bibliography{cite}
\begin{appendix}
\onecolumn
\section{Detailed Calculations}
\subsection{Calculation of the power spectrum}
Here we perform the calculation for the observed power spectrum $P^{\text{obs}}(\b \ell)$. For this, we assume an infinitely large field in order to perform our integration over $\mathbb{R}^2$. In reality, finite field effects would play a role. We begin with the calculation of the correlation for the Fourier transformed shear:
\begin{align}
& \la \gammaoh(\b \ell) \gammaoh {}^*(\b \ell')\ra\nonumber\\
 &\qquad = \la\int\text{d}^2 \theta\int\text{d}^2 \theta'\,W(\b \theta)W(\b \theta')\gamma(\b \theta)\gamma^*(\b \theta')\exp(\i \b \ell\b \theta-\i \b \ell'\b \theta')\ra\nonumber\\
 &\qquad = \la\int\text{d}^2 \theta\int\text{d}^2 \theta'\,W(\b \theta)W(\b \theta')\exp(\i \b \ell\b \theta-\i \b \ell'\b \theta') \int \frac{\text{d}^2 k}{(2\pi)^2}\int \frac{\text{d}^2 k'}{(2\pi)^2}\, \hat{\gamma}(\b k)\hat{\gamma}^*(\b k')\exp(-\i \b k\b \theta+\i \b k'\b \theta')\ra \nonumber\\
&\qquad = \la \int \text{d}^2 \theta \int \text{d}^2 \theta' \int \frac{\text{d}^2 k}{(2\pi)^2} \int \frac{\text{d}^2 k'}{(2\pi)^2}\, P(\b k)(2\pi)^2 \delta(\b k-\b k') \exp[\i (\b \ell\b \theta-\b \ell'\b \theta'-\b k\b \theta+\b k'\b \theta')]W(\b \theta)W(\b \theta')\ra \nonumber\\
  &\qquad = \la \int \frac{\text{d}^2 k}{(2\pi)^2} \, P(\b k) \int \text{d}^2 \theta\, W(\b \theta)\exp[\i\b \theta(\b \ell-\b k)]\int \text{d}^2 \theta'\, W(\b \theta') \exp[-\i\b \theta'(\b \ell'-\b k)] \ra \nonumber\\
  &\qquad = \la \int\frac{\text{d}^2 k}{(2\pi)^2} \, P(\b k)\widehat{W}(\b \ell-\b k)\widehat{W}^* (\b \ell'-\b k)\ra
\end{align}
It is important to keep in mind, that the ensemble averages of the weight function are independent of the ensemble averages of the shear values, meaning $\la W(\b \theta)\gamma(\b \theta)\ra = \la W(\b \theta)\ra \la \gamma(\b \theta)\ra$. We can define $W(\b \theta)=1+w(\b\theta)$ with $\la w(\b \theta)\ra = 0$, which leads to the expession
\begin{align}
& \la \gammaoh(\b \ell) \gammaoh {}^*(\b \ell')\ra \nonumber\\
 &\qquad = \la \int\frac{\text{d}^2 k}{(2\pi)^2} \, P(\b k) \left\{ (2\pi)^4\delta(\b \ell-\b k)\delta(\b \ell'-\b k)+(2\pi)^2\big[ \hat{w}(\b \ell-\b k)\delta(\b \ell'-\b k) + \hat{w}^*(\b \ell'-\b k)\delta(\b \ell-\b k)\big] + \hat{w}(\b \ell-\b k)\hat{w}(\b \ell'-\b k) \right\} \right> \nonumber\\
 &\qquad =  (2\pi)^2\delta(\b \ell-\b \ell')P(\b \ell) + \left[ \la \hat{w}(\b \ell-\b \ell')\ra P(\b \ell')+\la \hat{w}^*(\b \ell'-\b \ell)\ra P(\b \ell)\right] + \la \int \frac{\text{d}^2 k}{(2\pi)^2} \, \hat{w}(\b \ell-\b k)\hat{w}^*(\b \ell'-\b k)P(\b k)\ra \nonumber\\
& \qquad = (2\pi)^2\delta(\b \ell-\b \ell')P(\b \ell) + \la \int \frac{\text{d}^2 k}{(2\pi)^2}\, \hat{w}(\b \ell-\b k)\hat{w}^*(\b \ell'-\b k)P(\b k)\ra \, ,
\label{eq:pobs1}
\end{align}
where in the final step we have used that the average $\la \hat{w}(\b \ell)\ra$ vanishes.
Up until now, we have not specified our weight-function $w$. We parametrize it as \begin{equation}
w(\b \theta) = \sum_{\b \alpha \in \mathbb{Z}^2} w_{\b \alpha} \Xi(\b \theta-L\b \alpha)\text{ , with the box-function } \Xi(\b \theta) = \begin{cases}
1 \qquad \b \theta\in \left[-\frac{L}{2},\frac{L}{2}\right]^2 \\
0 \qquad \text{else}
\end{cases}.
\end{equation}
Here, the $w_{\b \alpha}$ are random variables, drawn from the random distribution describing the survey depths. For the Fourier transform we compute: \begin{equation}
\hat{w}(\b \ell) = \sum_{\b \alpha \in \mathbb{Z}^2} w_{\b \alpha} \exp(-\i L \b \ell \cdot\b \alpha) \widehat{\Xi}(\b \ell)\, ,
\end{equation}
where
\begin{equation}
\widehat{\Xi}(\b\ell) = \frac{4\sin\left(\frac{L\ell_1}{2}\right)\sin\left(\frac{L\ell_2}{2}\right)}{\ell_1\ell_2}\, ,
\label{eq:sinc}
\end{equation}
is a 2-dimensional sinc function.
Assuming an uncorrelated weight-distribution $\left(\la w_{\b \alpha} w_{\b \beta}\ra = 0\text{ for }\b \alpha\neq\b \beta\right)$ and setting $\la w^2\ra \equiv \la w_{\b \alpha}^2\ra$ for each $\b \alpha$, we get
\begin{align}
&\la \int \frac{\text{d}^2 k}{(2\pi)^2}\, \hat{w}(\b \ell-\b k)\hat{w}^*(\b \ell'-\b k)P(\b k)\ra \nonumber\\
&\qquad = \la \int \frac{\text{d}^2 k}{(2\pi)^2} \sum_{\b \alpha,\b \beta}w_{\b \alpha}w_{\b \beta} \exp[-\i L(\b \ell - \b k)\cdot\b \alpha]\, \widehat{\Xi}(\b \ell - \b k) \exp[\i L(\b \ell' - \b k)\cdot\b \beta]\, \widehat{\Xi}^*(\b \ell' - \b k)P(\b k)\ra \nonumber\\
&\qquad = \int \frac{\text{d}^2 k}{(2\pi)^2} \sum_{\b \alpha} \la w^2\ra \exp[-\i L(\b \ell - \b k)\cdot\b \alpha + \i L(\b \ell' - \b k) \cdot\b \alpha]\, \widehat{\Xi}(\b \ell - \b k)\widehat{\Xi}^*(\b \ell' - \b k)P(\b k) \, .
\end{align}
Using this result, we can obtain the observed power spectrum \begin{equation}
P^{\rm{obs}}(\b\ell) = \frac{1}{(2\pi)^2}\int\d^2\ell' \la \gammaoh(\b \ell) \gammaoh {}^*(\b \ell')\ra\, ,
\end{equation}
by performing the $\b \ell'$-integration in \eqref{eq:pobs1}:
\begin{align}
P^{\text{obs}}(\b \ell) = & P(\b \ell)+\int \frac{\text{d}^2\b \ell'}{(2\pi)^2} \int\frac{\text{d}^2\b k}{(2\pi)^2}\sum_{\b \alpha}\la w^2\ra \exp[-\i L (\b \ell-\b k)\cdot\b \alpha+\i L(\b \ell'-\b k) \cdot \b \alpha]\, \widehat{\Xi}(\b \ell-\b k)\widehat{\Xi}(\b \ell' - \b k) P(\b k) \nonumber\\
= & P(\b \ell)+ \int\frac{\text{d}^2\b k}{(2\pi)^2}\sum_{\b \alpha}\la w^2\ra \exp[-\i L (\b \ell-\b k) \cdot \b \alpha]\, \widehat{\Xi}(\b \ell - \b k)P(k) \int\frac{\text{d}^2\b \ell}{(2\pi)^2}\, \widehat{\Xi}^*(\b \ell'-\b k)\exp[\i L (\b \ell'- \b k)\cdot\b \alpha] \nonumber\\
 = & P(\b \ell) + \la w^2\ra \int\frac{\text{d}^2\b k}{(2\pi)^2}\, \widehat{\Xi}(\b \ell-\b k)P(\b k) \sum_{\b \alpha}\exp[-\i L (\b \ell - \b k) \cdot \b \alpha]\,\Xi(L\b \alpha) \nonumber\\
 = & P(\b \ell) + \la w^2\ra \int\frac{\text{d}^2\b k}{(2\pi)^2}\,\widehat{\Xi}(\b \ell-\b k)P(\b k)\, ,
\end{align}
which is a convolution of the power spectrum and the 2-dimensional sinc function \eqref{eq:sinc}. We note that due to the statistical inhomogeneity of the field, many usually adapted conventions fail. In particular, $\langle \gamma(\b \theta)\gamma^*(\b \theta')\rangle$ does not only depend on the separation vector $\b \theta'-\b \theta$, but also on the position $\b \theta$. For example, the Fourier transform of the observed power spectrum yields $\langle \gamma(\b 0)\gamma^*(\b\theta)\rangle$, but not the shear correlation function.
\label{sec:calc of PS}

\subsection{The function $E(\theta)$}
\label{sec:model_e}
{
\def\vec{\b}
\begin{figure}
    \sidecaption
    \centering
    \def\svgwidth{200pt}    
    \hspace*{1cm}
    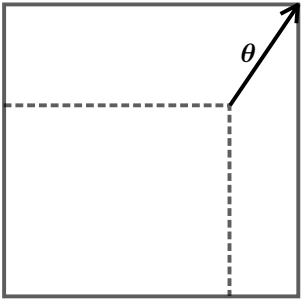  
    \hspace*{-2cm}
    \caption{Graphic representation on how to obtain the function $E(\b\theta)$. For a separation vector $\b\theta$, the dashed square represents the area of galaxies that have their partner in the same pointing.}
    \label{fig:explain_etheta}
\end{figure}
}
When computing the shear correlation between a pair of galaxies, it is of central importance whether or not those two galaxies lie in the same pointing. We want to model the probability that a pair of galaxies with separation $\b\theta$ lies in the same pointing by the function $E(\b\theta)$, which we will derive here:
Given one square field of length $L$ (in our case $L=60\arcmin$) and a separation vector $\b\theta = (\theta_1,\theta_2)$, without loss of generality we can assume $\theta_1,\theta_2\geq 0$. The dashed square in Fig.~\ref{fig:explain_etheta} represents all possible positions that the first galaxy can take, such that the second galaxy is still within the same pointing. The volume of this square equals \begin{equation}
V(|\b\theta|,\phi)  = \big[L-|\b\theta|\cos(\phi)\big]\,\big[L-|\b\theta|\sin(\phi)\big]\, ,
\end{equation} where $\phi$ represents the polar angle of the vector $\b\theta$. The function $E(\b\theta)$ then simply equals $V(|\b\theta|,\phi)/L^2$. To exclude negative volumes (which could occur when $|\b\theta|>1$ holds), we need to add the Heaviside step function $\mathcal{H}$:
\begin{equation}
E(\b\theta)  = \left[1-\frac{|\b\theta|}{L}\cos(\phi)\right]\,\left[1-\frac{|\b\theta|}{L}\sin(\phi)\right]\, \mathcal{H}\left[1-\frac{|\b\theta|}{L}\cos(\phi)\right]\,\mathcal{H}\left[1-\frac{|\b\theta|}{L}\sin(\phi)\right]\, .
\label{eq:eoftheta1}
\end{equation} 
As $E(\b\theta)$ is not isotropic, in order to obtain the function $E(\theta) = E(|\b\theta|)$, we need to calculate the azimuthal average of Eq.~\eqref{eq:eoftheta1} over all angles $\phi$. While the case $\theta_1,\theta_2\geq 0$ certainly does not hold for all angles $\phi$, we can omit the other cases by making use of the symmetry of the problem.
\begin{align}
E(\theta) = & \frac{4}{2\pi}\int_0^{\frac{\pi}{2}}\d\phi\, E(\b\theta) = \frac{2}{\pi}\begin{cases}
\int_0^{\frac{\pi}{2}} \d\phi\, \left[1-\frac{|\b\theta|}{L}\cos(\phi)\right]\,\left[1-\frac{|\b\theta|}{L}\sin(\phi)\right]\, ,  & |\b\theta|  \leq L \\[10pt]
\int_{\arccos(L/|\b\theta|)}^{\arcsin(L/|\b\theta|)} \d\phi\, \left[1-\frac{|\b\theta|}{L}\cos(\phi)\right]\,\left[1-\frac{|\b\theta|}{L}\sin(\phi)\right] \, , \quad   & L \leq |\b\theta| \leq \sqrt{2}L \\[10pt]
0\, , &\sqrt{2}L \leq\theta
\end{cases} \nonumber\\[10pt]
 = & \begin{cases}
\frac{1}{L^2 \pi}\left[L^2\pi - (4L-\theta) \theta\right]\, ,  & \theta \leq L \\[10pt]
\frac{2}{\pi}\,\left[4\sqrt{\frac{\theta^2}{L^2}-1} -1 - \frac{\theta^2}{2L^2} - \arccos\left(\frac{L}{\theta}\right) + \arcsin\left(\frac{L}{\theta}\right)\right]\, ,  & L  \leq \theta \leq \sqrt{2}L \\[10pt]
0\, ,  & \sqrt{2}L \leq \theta
\end{cases}\, .
\end{align}

\subsection{Calculation of the shear correlation functions}
\label{sec:calc of xipm}
Given a set of galaxies, we calculate the shear correlation function $\xi_+^{ij}$ via \begin{equation}
\xi^{ij}_+(\theta) = \frac{\sum_{a,b}w_a^iw_b^j\epsilon_a^i\epsilon_b^{j*}\Delta(|\b\theta_a^i-\b\theta_b^i|)}{\sum_{a,b}w_a^iw_b^j\Delta(|\b\theta_a^i-\b\theta_b^i|)}\, .
\label{eq:xip_from_observations}
\end{equation}
Here, $w$ represents the lensing weight of the galaxy, whereas $\epsilon$ is its (complex) ellipticity and $\b \theta$ its position on the sky. We have defined the function $\Delta$ as \begin{equation}
\Delta(|\b\theta_a^i-\b\theta_b^i|) = \begin{cases}
1, \,\, & |\b\theta_a^i-\b\theta_b^j| \in [\theta,\theta+{\rm d}\theta] \\
0, & \text{ else}
\end{cases}\, ,
\end{equation}
where we assume ${\rm d}\theta \ll \theta$. We define $\mathcal{N}$ as the number of pointings in the survey and $F_k^i$ as the set of galaxies in pointing $k$ and tomographic redshift bin $i$. The numerator in Eq.~\eqref{eq:xip_from_observations} then transforms to: \begin{align}
& \sum_{k,\ell=1}^\mathcal{N} \sum_{a\in F_k^i}\sum_{b\in F_{\ell}^j} w_a^iw_b^j\epsilon_a^i\epsilon_b^{j*} \Delta(|\b\theta_a^i-\b\theta_b^i|) \nonumber\\
 = & \sum_{k=1}^\mathcal{N}\sum_{a\in F_k^i}w_a^i \sum_{\ell=1}^\mathcal{N} \sum_{b\in F_{\ell}^j} w_b^j \Delta(|\b\theta_a^i-\b\theta_b^i|)\, \epsilon_a^i\epsilon_b^{j*} \nonumber\\
  = & \sum_{k=1}^\mathcal{N}\sum_{a\in F_k^i}w_a^i \left[\sum_{b\in F_{k}^j} w_b^j \Delta(|\b\theta_a^i-\b\theta_b^i|)\, \epsilon_a^i\epsilon_b^{j*} + \sum_{\ell\neq k} \sum_{b\in F_{\ell}^j} w_b^j \Delta(|\b\theta_a^i-\b\theta_b^i|)\, \epsilon_a^i\epsilon_b^{j*}\right] \, .
\end{align}
When we denote the probability that pointing $k$ is of quantile $m$ by $\mathcal{P}_m^k$ and assume that the product $\epsilon_a^i\epsilon_b^{j*}$ always equals its expectation value, we can set the numerator as \begin{equation}
\sum_{k=1}^\mathcal{N}\sum_{a\in F_k^i}w_a^i \sum_m \mathcal{P}_m^k \left[\overbrace{\sum_{b\in F_{k}^j} w_b^j \Delta(|\b\theta_a^i-\b\theta_b^i|)}^{\rm{(\ref{eq:numerator1}.a)}}\,  \xi_{+,mm}^{ij}(\theta) + \overbrace{\sum_{\ell\neq k} \sum_{b\in F_{\ell}^j} w_b^j \Delta(|\b\theta_a^i-\b\theta_b^i|)}^{\rm{(\ref{eq:numerator1}.b)}} \sum_n \mathcal{P}_n^{\ell} \xi_{+,mn}^{ij}(\theta)\right] \, .
\label{eq:numerator1}
\end{equation}
The term (\ref{eq:numerator1}.a) denotes all galaxies that lie within distance interval $[\theta,\theta+{\rm d}\theta]$ of galaxy $a$, and are in the same pointing as galaxy $a$. This term is equal to the (weighted) number density of galaxies in the pointing multiplied by $2\pi\theta\, {\rm d}\theta\, E(\theta)$. \\
The term (\ref{eq:numerator1}.b) denotes all galaxies within distance interval $[\theta,\theta+{\rm d}\theta]$ of galaxy $a$, that are \textit{not} in the same pointing as galaxy $a$. This is equal to the number density of galaxies in the respective pointings multiplied by $2\pi\theta\, {\rm d}\theta\, [1-E(\theta)]$. 
 
If we assume that said number density in a pointing is equal to the number density in the quantile it belongs to, $\hat{n}_n^j$, and set $\mathcal{P}_n^\ell=1/10$, the numerator becomes \begin{align}
\sum_{k=1}^\mathcal{N} \sum_{a\in F_k^i} w_a^i \sum_m P^k_m \left[2\pi\theta\,\d\theta\,E(\theta) \hat{n}_m^j \xi_{+,mm}^{ij}(\theta) + 2\pi\theta\,\d\theta\,\frac{1-E(\theta)}{10}\, \sum_n \hat{n}_n^j \xi_{+,mn}^{ij}(\theta) \right] \, .
\end{align}
The term $\sum_{a\in F_k^i} w_a^i$ denotes the number of galaxies in pointing $k$, which we set as the number density of galaxies in the respective quantile multiplied with the area $A$ of the pointing. Applying this and setting $\mathcal{P}_m^k=1/10$, the numerator reads 
\begin{align}
\frac{2\pi\theta\,\d\theta}{10} \sum_{k=1}^\mathcal{N} \sum_m \hat{n}_m^i A &\left[ E(\theta) N_m^j \xi_{+,mm}^{ij}(\theta) + \frac{1-E(\theta)}{10}\sum_n \hat{n}_n^j\xi_{+,mn}^{ij}(\theta)\right] \nonumber\\
 =  \, \frac{2\pi\theta\,\d\theta \, NA}{10}  \sum_m \hat{n}_m^i &\left[ E(\theta) \hat{n}_m^j \xi_{+,mm}^{ij}(\theta) + \frac{1-E(\theta)}{10}\sum_n \hat{n}_n^j\xi_{+,mn}^{ij}(\theta)\right] \, .
\end{align}
 The same line of argumentation can be applied to the denominator, which then reads: \begin{equation}
\frac{2\pi\theta\,\d\theta \, NA}{10}  \sum_m \hat{n}_m^i \left[ E(\theta) \hat{n}_m^j + \frac{1-E(\theta)}{10}\sum_n \hat{n}_n^j\right]\, .
 \end{equation}
Taking the ratio of the two quantities, and setting $N_n^i = A\hat{n}_n^i$, we see that Equations \eqref{eq:xip_from_observations} and \eqref{eq:correctionfunction1} are the same.
\section{Finite field effects}
\label{sec:expand_eoftheta}
\begin{figure}
    \sidecaption
    \centering
    \def\svgwidth{200pt}    
    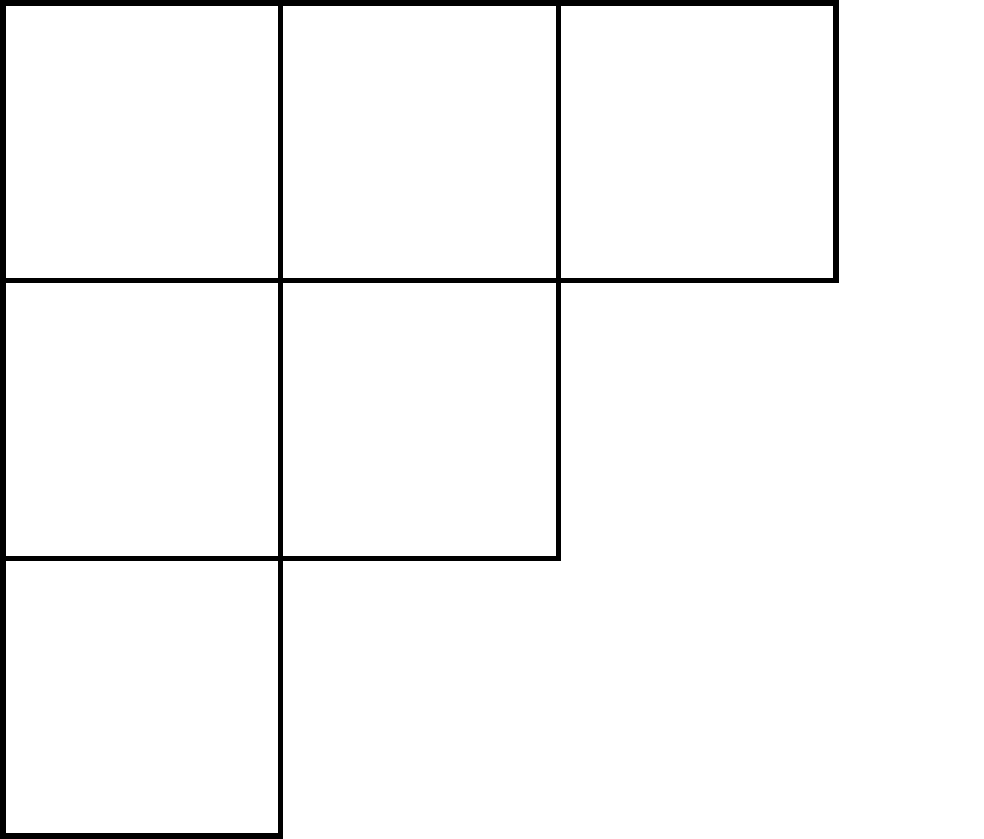  
    \caption{Graphic representation of the definitions of $E_{ab}(\theta)$. When the first galaxy is in the bottom left pointing, the probability to find the second galaxy in a pointing of distance $(a,b)$ is $E_{ab}(\theta)$.}
    \label{fig:expand_etheta}
\end{figure}
In this appendix, we outline how to calculate the correction of the correlation functions for a finite survey with a potentially correlated distribution of depth between pointings. Essentially, this boils down to the calculation of $\mathcal{P}_{mn}^{ij}(\theta)$ from Eq.~\eqref{eq:def_xiobs}. We calculate this weighting by the geometrical probability that a pair of galaxies of separation $\theta$ is of quantiles $m$ and $n$, $\mathcal{P}(m,n|\theta)$, weighted by the respective number of galaxies in the quantiles $N_m^i,N_n^j$: \begin{equation}
\mathcal{P}_{mn}^{ij}(\theta) = N_m^iN_n^j\mathcal{P}(m,n|\theta)\, .
\label{eq:pmnij1}
\end{equation}
 At first, we define functions $E_{ab}(\b\theta)$ as the probabilty that a galaxy pair of separation $\b\theta$ is in pointings of distance $(a,b)$. This situation is depicted in Fig.~\ref{fig:expand_etheta}. Due to symmetry, for the azimuthal average of the functions, $E_{ab}(\theta) = E_{-ab}(\theta) = E_{ba}(\theta)$ holds for all combinations of $a$ and $b$. We note that $E_{00}(\theta)=E(\theta)$ and $\sum_{a,b}E_{ab}(\theta)\equiv 1$.

Let $\mathcal{P}^*(m,n|a,b)$ denote the probability that two pointings of distance $(a,b)$ are of quantile $m$ and $n$ (which is directly calculable from a given survey footprint). 
Then the following equation holds: \begin{equation}
\mathcal{P}(m,n|\theta) = \sum_{a,b} E_{ab}(\theta)\mathcal{P}^*(m,n|a,b)\, .
\label{eq:pmntheta}
\end{equation}
Note that the expectation value of $\mathcal{P}^*(m,n|a,b)$ for uncorrelated distributions is \begin{equation}
\la \mathcal{P}^*(m,n|a,b)\ra = \begin{cases}
0.1\,\delta_{mn},\qquad & \text{for }(a,b)=(0,0) \\
0.01, & \text{else}
\end{cases}\, ,
\label{eq:pmnabstar}
\end{equation}
where $\delta_{mn}$ denotes the Kronecker delta. Keeping in mind that \begin{equation}
\sum_{(a,b)\neq (0,0)} E_{ab}(\theta) = 1-E(\theta)\, ,
\end{equation}
we can use the expectation value \eqref{eq:pmnabstar} to calculate \eqref{eq:pmntheta} as a consistency check. In that case, we receive the same value for the coefficients in \eqref{eq:pmnij1} as we have in Eq. \eqref{eq:pmnij_uncorr} in Sec. \ref{sec:xipm_semianalytic} for the case of an infinite footprint and uncorrelated distribution of depth.

\begin{figure}
\centering
\sidecaption
$
\begin{array}{cc}
\def\svgwidth{100pt}
\hspace*{1cm}
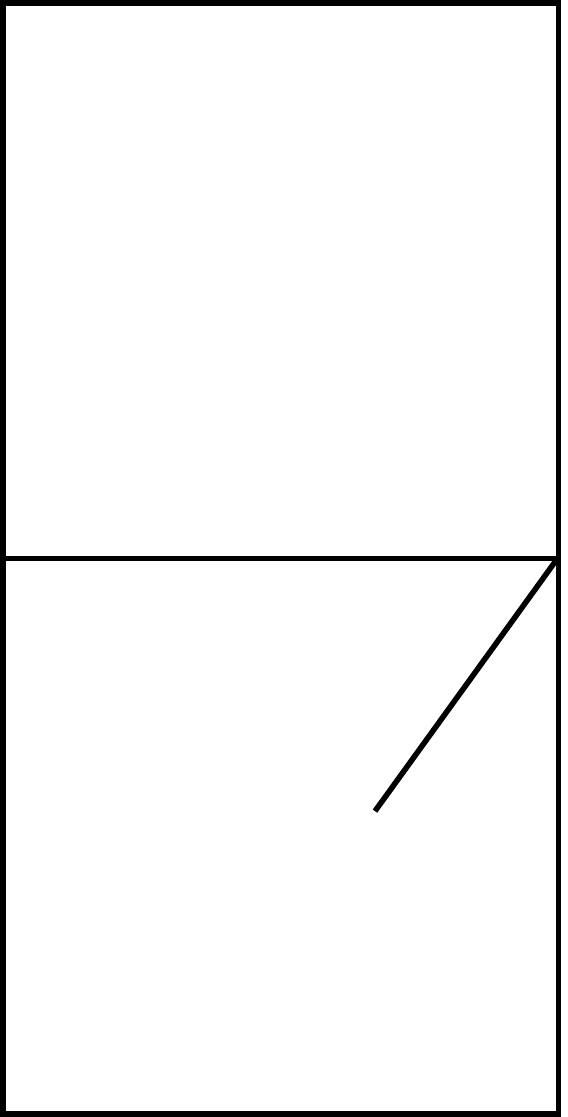&
\def\svgwidth{100pt}
\hspace*{1cm}
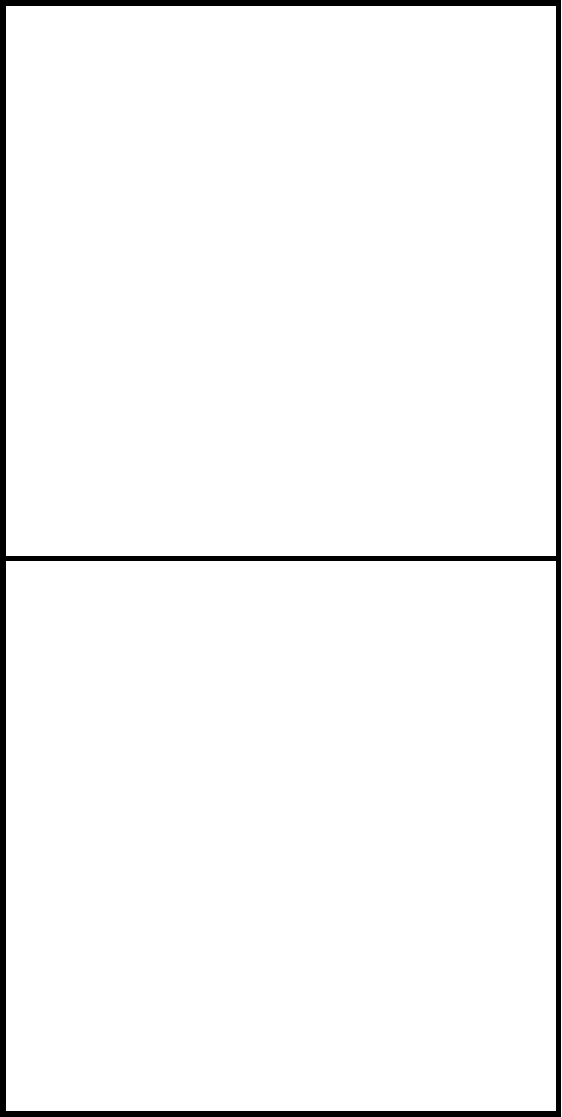
\end{array}$
\caption[How to calculate $E_{01}(\theta)$]{Representation of how to calculate $E_{01}(\theta)$ for different values of $\theta$. For $\theta\sin(\phi)<L$, as depicted in the left part, the volume of the dashed rectangle is $V(\theta,\phi)=\theta\sin(\phi)[L-\theta\cos(\phi)]$. For $\theta\sin(\phi)>L$, as depicted in the right part, the volume of the dashed rectangle is $V(\theta,\phi)=[2L-\theta\sin(\phi)]\, [L-\theta\cos(\phi)]$.}
\label{fig:e01theta}
\end{figure}

The $E_{ab}$ can all be calculated analytically, similar to our method in Sec. \ref{sec:model_e}. We again assume a selection of square fields with side length $L$, and later set $L=60\arcmin$ to adapt to the KV450 survey. As an example, for $E_{01}$ we have several possible situations, depicted in Fig. \ref{fig:e01theta}. Setting $E_{ab}(\b\theta) = V(\theta,\phi)/L^2$, we define \begin{align}
E_{01}^{(a)}(\b\theta) & \equiv \frac{\theta}{L}\sin(\phi)\left[1-\frac{\theta}{L}\cos(\phi)\right] \nonumber\\
E_{01}^{(b)}(\b\theta) & \equiv \left[2-\frac{\theta}{L}\sin(\phi)\right]\, \left[1-\frac{\theta}{L}\cos(\phi)\right]
\end{align}
With some geometric considerations, we compute:
{
\begingroup
\addtolength{\jot}{1em}
\begin{align}
E_{01}(\theta) = & \begin{cases}
\frac{1}{\pi}\int_0^{\frac{\pi}{2}} \d\phi\, E_{01}^{(a)}(\b\theta), & \frac{\theta}{L} < 1 \\[10pt]
 \frac{1}{\pi}  \left[\int_{\arccos(L/\theta)}^{\arcsin(L/\theta)}\d\phi\,E_{01}^{(a)}(\b\theta) + \int_{\arcsin(L/\theta)}^{\frac{\pi}{2}}\d\phi\, E_{01}^{(b)}(\b\theta) \right],  & 1 < \frac{\theta}{L} < \sqrt{2} \\[10pt]
 \frac{1}{\pi} \int_{\arccos(L/\theta)}^{\frac{\pi}{2}}\d\phi\, E_{01}^{(b)}(\b\theta), & \sqrt{2}<\frac{\theta}{L}<2 \\[10pt]
\frac{1}{\pi} \int_{\arccos(L/\theta)}^{\arcsin(2L/\theta)}\d\phi\, E_{01}^{(b)}(\b\theta), & 2<\frac{\theta}{L}<\sqrt{5} \\[10pt]
 0, & \sqrt{5}<\frac{\theta}{L}
\end{cases}\nonumber\\
 = & \begin{cases}
 \frac{(2L-\theta)\theta}{2\pi L^2}, & \frac{\theta}{L} < 1 \\[10pt]
 \frac{1}{\pi}\left[\frac{3}{2}- 2\frac{\theta}{L}+\frac{\theta^2}{L^2}+2\sqrt{\frac{\theta^2}{L^2}-1}+2\arcsin\left(\frac{L}{\theta}\right)\right], & 1<\frac{\theta}{L}<\sqrt{2} \\[10pt]
 \frac{1}{2\pi}\left[-1-4\frac{\theta}{L}+4\sqrt{\frac{\theta^2}{L^2}-1}+4\arccos\left(\frac{L}{\theta}\right)\right], & \sqrt{2} < \frac{\theta}{L} < 2 \\[10pt]
 \frac{1}{2\pi}\left[-5-\frac{\theta^2}{L^2}+2\sqrt{\frac{\theta^2}{L^2}-4}+4\sqrt{\frac{\theta^2}{L^2}-1}-4\arcsin\left(\frac{L}{\theta}\right)+4\arcsin\left(\frac{2L}{\theta}\right)\right], & 2 < \frac{\theta}{L} < \sqrt{5} \\[10pt]
 0, & \sqrt{5}<\frac{\theta}{L}
 \end{cases}\, .
\end{align}
\endgroup
}

\begin{figure}
\centering
\sidecaption
\hspace*{1cm}
\def\svgwidth{140pt}
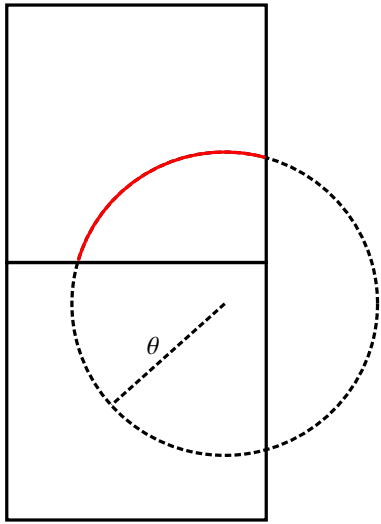
\caption{Visualization of the numerical computation for $E_{01}(\theta)$. For a circle of radius $\theta$, the length of the red arc divided by $2\pi$ represents the fraction of galaxies within the respective pointing. This value needs to be integrated for all possible centers of the circle in the pointing. That procedure is straightforward to expand for other $E_{ab}(\theta)$.}
\label{fig:eoftheta_sim}
\end{figure}

\begin{figure}[!htb]
\centering
\sidecaption
\includegraphics[width = 0.6\linewidth]{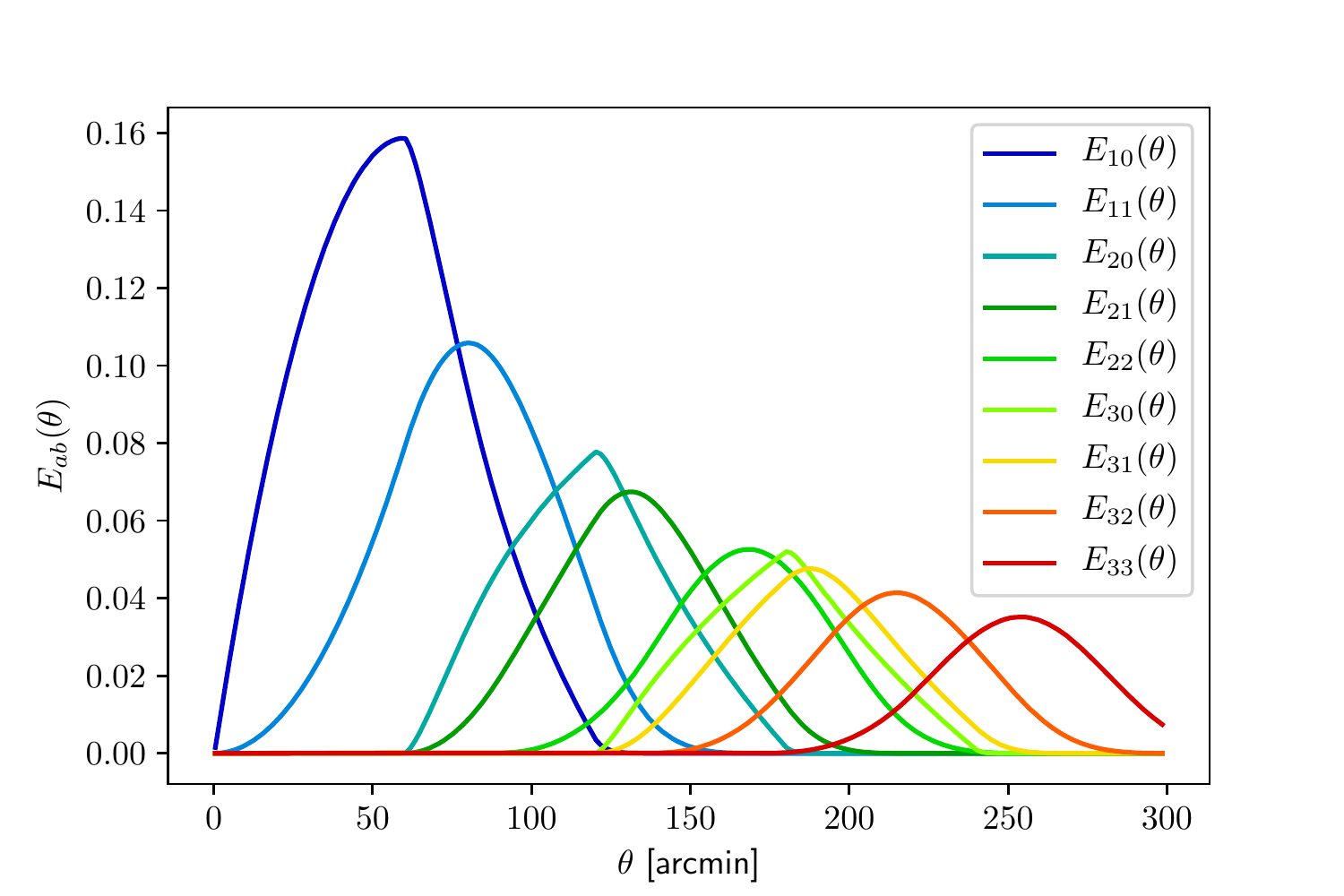}
\caption{The functions $E_{ab}(\theta)$ for the first few possible combinations.}
\label{fig:eab}
\end{figure}
Naturally, to calculate those functions for all possible combinations would be rather tedious, however they are simple to determine numerically (compare Fig.~\ref{fig:eoftheta_sim}). A plot of these functions can be found in Fig.~\ref{fig:eab}.

\begin{figure}[!htb]
\centering
\includegraphics[width=0.9\linewidth]{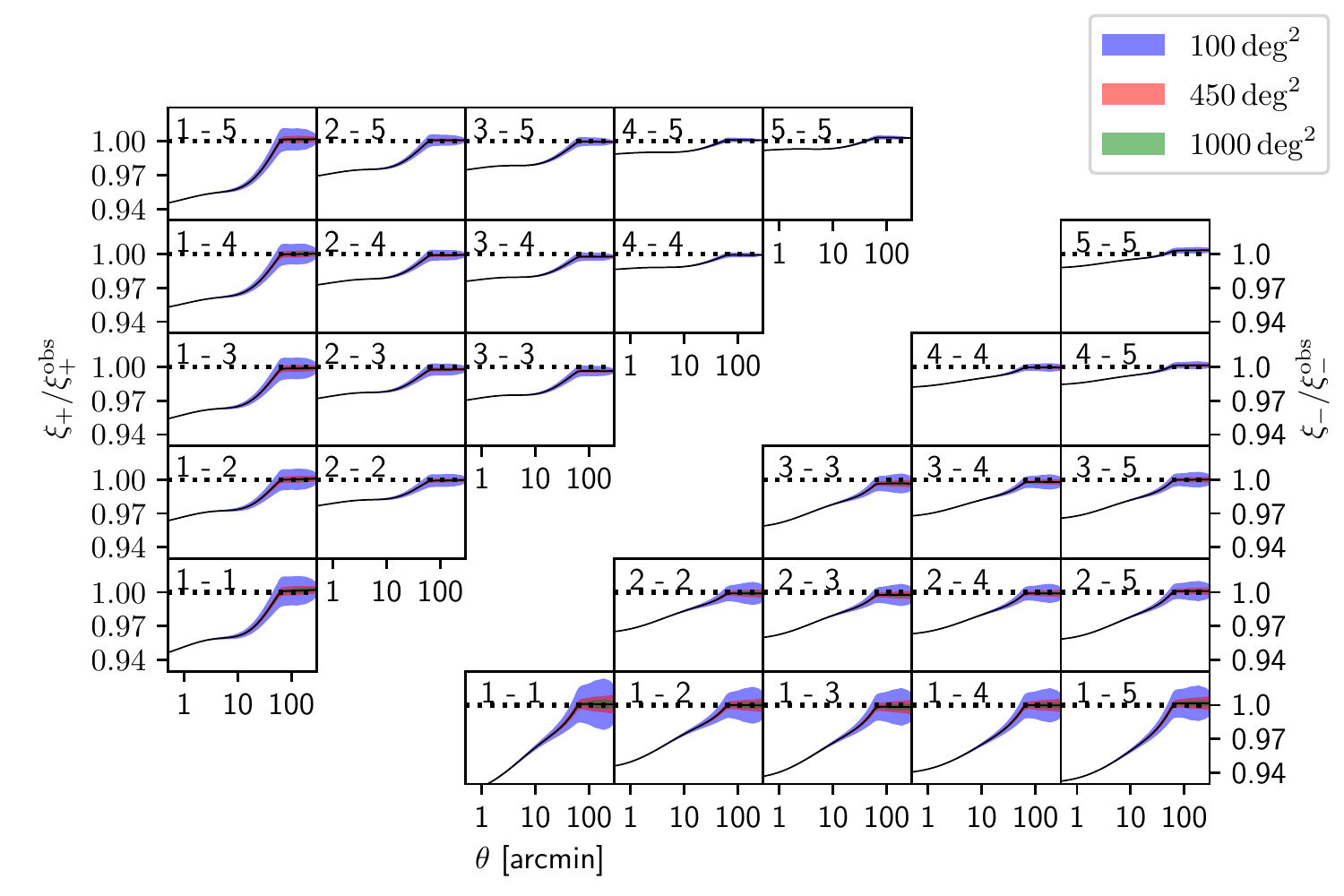}
\caption{$2\sigma$-contours of the corrections for the correlation functions for a $100\,\rm{deg}^2$ field (blue), a $450\,\rm{deg}^2$ field (red) and a $1000\,\rm{deg}^2$ field (green). As can be seen, the variance of the variation is small for a $450\,\rm{deg}^2$ field and barely noticeable for a $1000\,\rm{deg}^2$ field.}
\label{fig:finite_footprint}
\end{figure}

We sample several realizations of a random depth-distribution for a $100\,\rm{deg}^2$-field, a $450\,\rm{deg}^2$-field and a $1000\,\rm{deg}^2$-field. For each realization we extract the Function $\mathcal{P}^*(m,n|a,b)$ and, using Eq.~\eqref{eq:pmntheta}, calculate the ratio $\xi_\pm^{\rm{obs}}/\xi_\pm$. Afterwards, we compute the variance of these ratios. As can be seen from Fig. \ref{fig:finite_footprint}, the effect is quite significant for a $100\,\rm{deg}^2$-field, but almost negligible for a $1000\,\rm{deg}^2$-field. This leads to the assumption that both for the KV450 survey as well as for all next-generation cosmic shear surveys, finite field effects do not need to be accounted for. However, if the distribution of depth is correlated in the surveys, that might have a noticeable impact on the results.

\section{Additional Figures and Tables}
\begin{table}[!htb]
\centering
\caption{Limiting magnitudes for the ten quantiles}
\label{tab:maglim}
\begin{tabular}{r c}
\hline\hline
quantile & $r$-band depth \\
\hline
0 & 25.76 \\
10 & 26.06 \\
20 & 26.15 \\
30 & 26.19 \\
40 & 26.23 \\
50 & 26.27 \\
60 & 26.31 \\
70 & 26.34 \\
80 & 26.39 \\
90 & 26.44 \\
100 & 26.60 \\
\hline
\end{tabular}
\end{table}
\begin{figure}[!htb]
\centering
\includegraphics[width=0.9\linewidth]{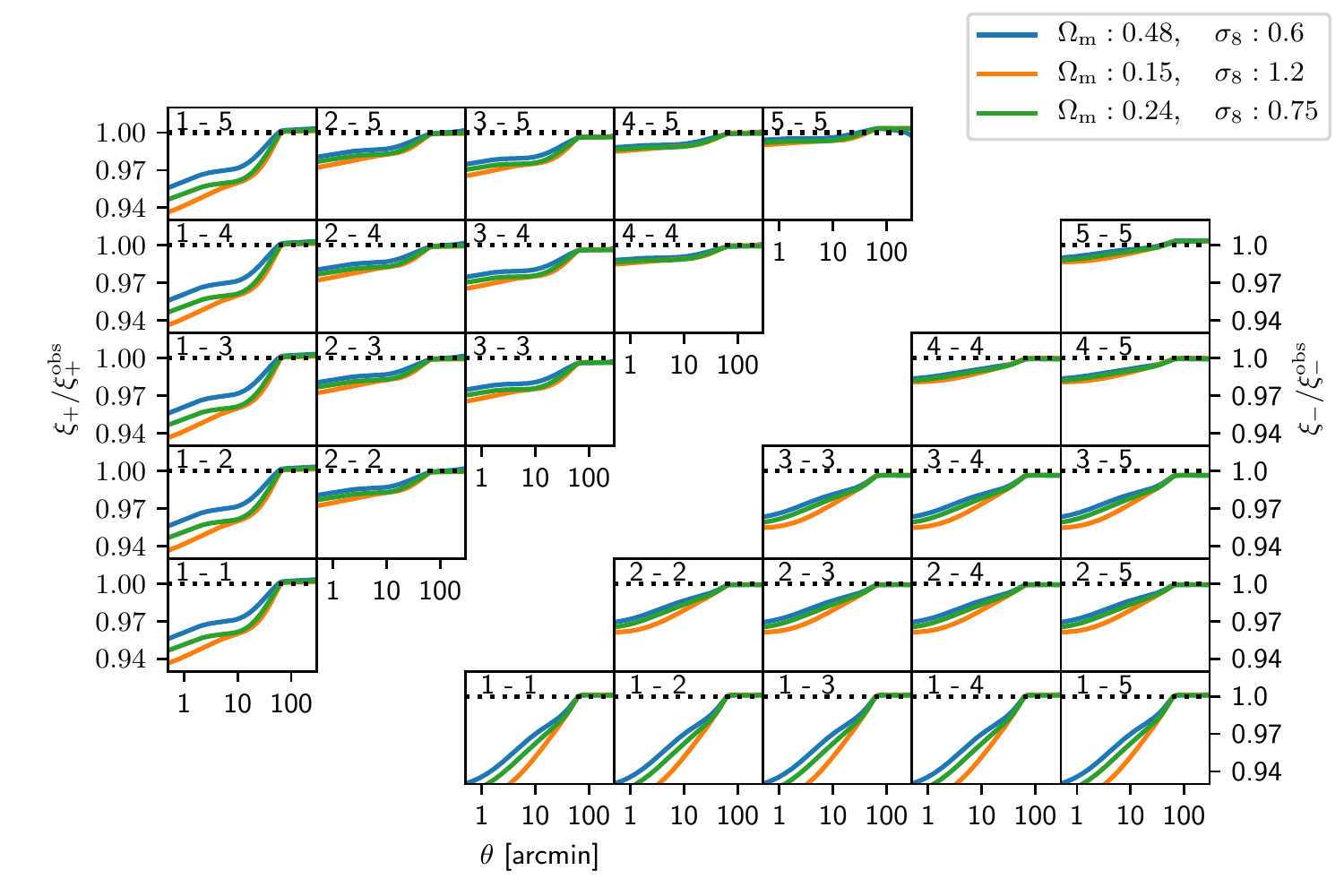}
\caption{Correction to the correlation functions in varying cosmologies. Depicted here are three flat sample cosmologies, where values within the 98\% CL of the KV450 survey were sampled.}
\label{fig:comparecosmo}
\end{figure}
\end{appendix}
\end{document}